\documentclass[11pt]{article} \hoffset=-15mm \voffset=-10mm
\usepackage{graphicx,amsmath,amssymb,epsf} \usepackage{latexsym,bm,
  slashed} \usepackage{xcolor} \definecolor{dark}{rgb}{0.10,0.2,0.3}
\definecolor{magenta}{rgb}{0.7,0.1,0.3}
\definecolor{purpure}{rgb}{0.5,0.15,0.3}
\usepackage[font=small,format=plain,labelfont=bf,up,textfont=it,up]{caption}
\usepackage{hyperref, cite} \hypersetup{colorlinks, citecolor=blue,
  filecolor=blue, linkcolor=magenta,
  urlcolor=purpure,hyperfootnotes=true,pdftex} 
  
\newcommand{\tr}{{\rm tr}}

\newcommand{\bx}{{\bm x}}

 \title{\bf
  \Large \bf \Large Spinor helicity methods in high-energy factorization: \\
efficient momentum-space calculations in \\
the
  Color Glass Condensate formalism} \author{ Alejandro~Ayala$^{1,2}$,
  Martin~Hentschinski$^{1,3}$, Jamal~Jalilian-Marian$^{4,5,6}$, \\
  Maria~Elena~Tejeda-Yeomans$^7$
  \bigskip \\
  { \normalsize $^1$Instituto de Ciencias Nucleares, Universidad
    Nacional Aut\'onoma de M\'exico,}
  \\
  {\normalsize Apartado Postal 70-543, Ciudad de M\'exico 04510, Mexico } \\
  {\normalsize $^2$Centre for Theoretical and Mathematical Physics,
    and Department
    of Physics}, \\
  {\normalsize University of Cape Town, Rondebosch 7700, South Africa}\\
  {\normalsize $^3$Departamento de Actur\'ia, Fisica y Matematicas,
    Universidad de las Americas - Puebla} \\ {\normalsize Santa
    Catarina Martir, 72820, Puebla, Mexico  } \\
  {\normalsize $^4$Department of Natural Sciences, Baruch College,
    CUNY,}
  \\
  {\normalsize 17 Lexington Avenue, New York, NY 10010, USA}\\
  {\normalsize $^5$CUNY Graduate Center, 365 Fifth Avenue, New York, NY 10016, USA}\\
  {\normalsize $^6$Centre de Physique Th\'eorique, \'Ecole
    Polytechnique, CNRS,}
  \\
  {\normalsize Universit\'e Paris-Saclay, 91128 Palaiseau, France}\\
  { \normalsize $^7$Departamento de F\'isica, Universidad de Sonora,
    Boulevard
    Luis Encinas J. y Rosales, } \\
  {\normalsize Colonia Centro, Hermosillo, Sonora 83000, Mexico} }

\begin{document}

\maketitle
\begin{abstract}

  We use the spinor helicity formalism to calculate the cross section
  for production of three partons of a given polarization in Deep
  Inelastic Scattering (DIS) off proton and nucleus targets at small
  Bjorken $x$. The target proton or nucleus is treated as a classical
  color field (shock wave) from which the produced partons scatter
  multiple times.  We reported our result for the final expression for
  the production cross section and studied the  azimuthal angular
  correlations of the produced partons in~\cite{ahjt-1}. Here we
  provide the full details of the calculation of the production cross
  section using the spinor helicity methods.

\end{abstract}

\section{Introduction}
\label{sec:introduction}

The Color Glass Condensate (CGC) formalism (see~\cite{gijv} for a
review) is an effective field theory approach to QCD at small $x$
where the gluon density in a proton or nucleus is expected to be so
large that the standard QCD-improved parton model must break down. In
this limit it is more appropriate to treat the hadron or nucleus as a
coherent color field rather than a collection of incoherent and
individual partons. At the classical level, the CGC generalizes
scattering via exchange of a single gluon to multiple gluon exchanges
(high energy factorization). By including quantum loop effects, it
resums the large logs of Bjorken $x$~\cite{jimwlk} (or equivalently
energy) as opposed to the leading twist collinear factorization
formalism which resums large logs of the hard scale $Q^2$. The Color
Glass Condensate formalism is the appropriate one to use in the
kinematic limit where $\log 1/x \gg \log Q^2$. This is typically the
case for particle production in the low to intermediate $p_t$ regime
and in the forward rapidity region~\cite{forward} where one probes the
small $x$ components of the target wave function.\\

The Color Glass Condensate formalism has been applied to various high
energy processes. A first class of observables is associated with the
forward hadron production at RHIC and the LHC (see
\cite{N.Cartiglia:2015gve} for a compilation of LHC forward physics
observables and \cite{predic} for a compilation of the CGC-motivated
phenomenology of particle production). While there are strong hints
for the presence of gluon saturation effects in the observed
suppression of forward hadron production cross section at RHIC and the
LHC~\cite{forward-single}, the effect is far from being universally
accepted as coming only from saturation. Models which modify the
parton distribution functions of nuclei and add energy loss effects
have also been used to fit the data~\cite{noncgc-single}. Azimuthal
angular correlations of produced di-hadrons have also been measured
(in deuteron-gold collisions) and show a strong decorrelation of the 
azimuthal angle between the two produced hadrons in the forward rapidity 
region as predicted in the CGC formalism~\cite{cm-double}. The away side peak 
is observed to be broader which can be understood to be due to multiple
scattering. Furthermore, the magnitude of the peak is reduced which is
due to the small $x$ evolution of the target wave function which
causes \lq\lq nuclear shadowing\rq\rq. Various angular correlations
have also been suggested as possible probes of gluon saturation in the
target~\cite{forward-double,JalilianMarian:2004da}. They all exhibit
the same qualitative features in the CGC formalism which will be
further probed by the current experiments at the LHC and in the proposed
Electron-Ion Collider~\cite{dis-double}.\\

While much can be learned from hadron-hadron collisions, observables
generally suffer from contaminations due to soft and collinear
radiation. A much cleaner environment -- both experimentally and
theoretically -- is, on the other hand, provided by photon induced
processes.  While much has been learned from the study of proton
structure functions measured at the HERA experiment, ultra-peripheral
collisions (UPCs) of protons on protons and nuclei on protons at the
LHC, allow to study the currently most energetic photon-proton
collisions, where either a large nucleus or a proton serve as the
photon source. Among the most prominent observables explored so far is
exclusive photo-production process of vector mesons such as the
$J/\Psi$ \cite{TheALICE:2014dwa,Aaij:2013jxj}. Such processes are of
particular interest since they allow to probe the gluon distribution
in the proton down to very small values of Bjorken $x \sim 10^{-5}$,
see \cite{Armesto:2014sma} for CGC-related studies. While such CGC
studies provide an excellent description of the energy dependence of
the exclusive $J/\Psi$ photo-production cross-section, the same energy
dependence has also an excellent description in terms of (dilute)
next-to-leading order BFKL evolution (see {\it e.g.}
\cite{Bautista:2016xnp}) which, unlike the CGC-formalism, does not
include corrections due to high densities.  From a theoretical point
of view the difficulty to distinguish low $x$ evolution without (BFKL)
and with (BK, JIMWLK) is related to the type of correlators probed by
such observables: both proton structure functions and the amplitude
for exclusive photo-production of vector mesons are directly
proportional to the two-point correlator of two Wilson lines, which
itself can be related to an unintegrated gluon density. The
description of such observables is therefore at first identical for
low (BFKL) and high (BK, JIMWLK) densities and the presence of high
density effects only manifests itself through the particular form of
low $x$ evolution. Finding definite evidence for saturation effects
requires therefore evolution of the observables far into the low $x$
region, well beyond the kinematical reach of current collider
experiments.
 \\

 In~\cite{ahjt-1} we proposed triple jet/hadron production in DIS as
 an excellent probe of gluon saturation dynamics in a hadron or
 nucleus.  Unlike the observables discussed above, the theoretical
 description of observables with multiple final states involves higher
 order correlators of Wilson lines in the target. As a consequence
 high density effects manifest themselves directly at the level of the
 observable without the need to invoke low $x$ evolution. In
 particular there are substantial differences between a BFKL/low
 density and a CGC/high density description. While the measurement of
 such high multiplicity final states is much more cumbersome
 experimentally, such observables provide important complementary
 information to inclusive observables, and therefore allow to pin down
 the valid description of perturbative QCD in the low $x$ region. The
 case of three jet/hadron production is then of particular interest,
 since such an observable provides two relative angles rather than one
 in the case of two-particle correlations, so there is an additional
 knob to turn. Furthermore there are two away side peaks, each of
 which will sensitively depend on saturation dynamics and small $x$
 evolution. We showed our final expression for the triple differential
 cross section in~\cite{ahjt-1} and used it to study, quantitatively,
 the angular dependence of the cross section in certain
 kinematics. The present paper is dedicated to a presentation of the
 full details of the calculation, employing spinor helicity formalism
 (for introductory reviews see~\cite{dixon}) which leads to a major
 simplification of the calculation. For another recent application of
 the spinor helicity formalism in the framework of light-cone wave
 function see \cite{Lappi:2016oup}; for a related calculation of 3
 parton impact factor in DIS in the shock wave framework
 see also \cite{Beuf:2011xd,Boussarie:2016ogo}; see also the
 calculation of next-to-leading corrections to inclusive DIS of \cite{Balitsky:2010ze,Beuf:2016wdz}.\\

 This paper is organized in the following way: we write down the
 formal expressions for the production amplitude in Section
 \ref{sec:ampl-moment} and proceed to study the space-time structure
 of the diagrams in Section \ref{sec:cut-diagr-accord} and
 \ref{sec:further-reduction}. Various cuts of the diagrams are
 considered and shown to lead to vanishing of some diagrams due to the
 \lq\lq wrong cut\rq\rq. This allows us to re-write the expressions
 for the amplitude in a more compact way. We then give an overview of
 the spinor helicity formalism in Section \ref{sec:helic-spin-meth}
 and proceed to apply it to the process considered. Major
 simplifications occur when one considers scattering and production of
 partons of a fixed helicity due to helicity conservation. We
 summarize in Section \ref{sec:summary} and, using crossing symmetry,
 point out the connections between our results and Multi-Parton
 Scattering (MPI) in the projectile, in photon-jet or photon
 production in forward rapidity proton-nucleus collisions.

\section{CGC amplitudes in momentum space}
\label{sec:ampl-moment}

\begin{figure}[t]
  \centering
  \includegraphics[width=.5\textwidth]{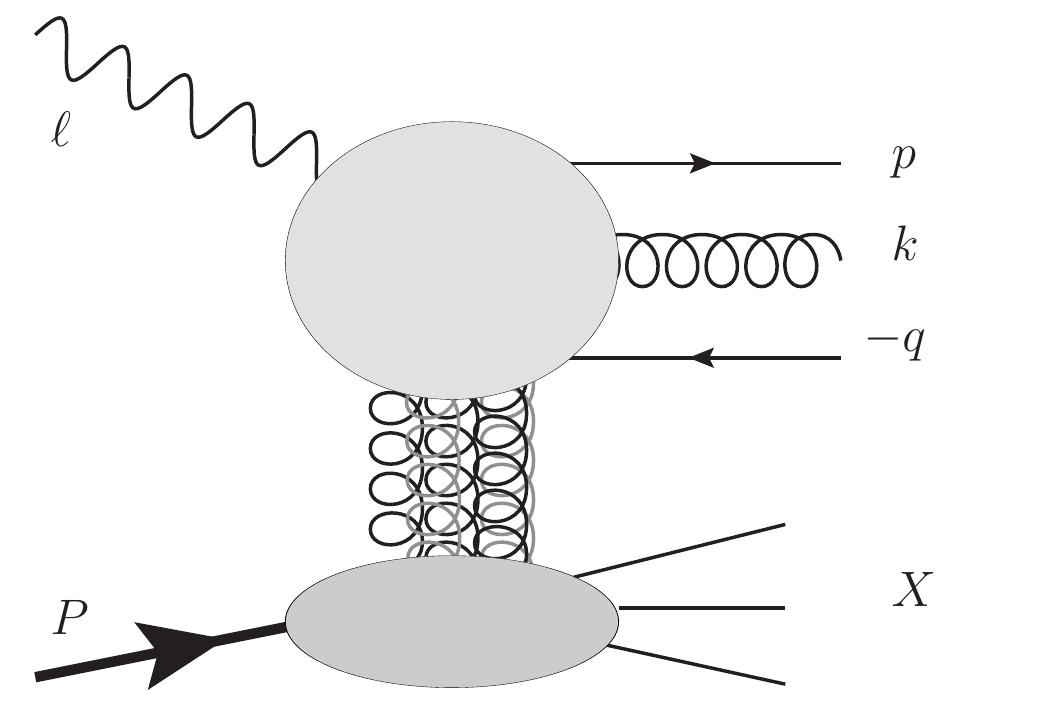}
  \caption{\it}
  \label{fig:process}
\end{figure}

\begin{figure}[t]
  \centering
  \includegraphics[width=.4\textwidth]{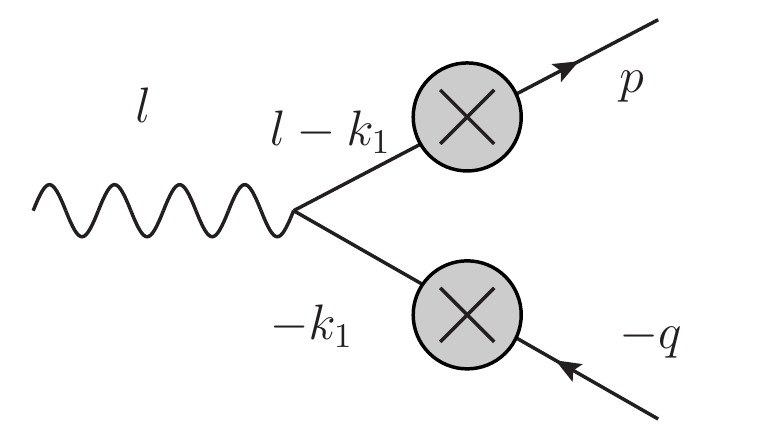}
\includegraphics[width=.4\textwidth]{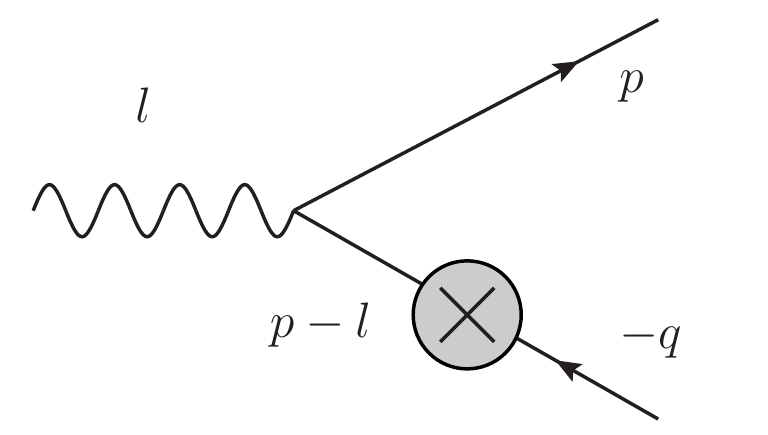}
  \caption{{\it Left}: Tree diagram with 2 insertions of the vertices Eqs.~\eqref{eq:quarkinteraction} and \eqref{eq:gluoninteraction}: the internal momentum $k_1$ is integrated over like a loop momenta {\it i.e.} with $\int d^4 k_1/(2 \pi)^4$. {\it Right}: Tree diagram with 1 insertion the vertices Eqs.~\eqref{eq:quarkinteraction} and \eqref{eq:gluoninteraction}: all momenta are fixed by external momenta }
  \label{fig:example_mom_fdiag}
\end{figure}

We consider production of three partons in DIS  as depicted in Fig.~\ref{fig:process},
\begin{align}
  \label{eq:process}
  \gamma^*(l) + \text{target}\, (P) & \to q(p) + \bar{q}(q) + g(k) + X,
\end{align}
where the target can be a proton or nucleus and where $Q^2 = - l^2$
denotes the virtuality of the photon. We will study the process in the
limit of high center of mass energy $\sqrt{s} \to \infty$ with
$s= (l + P)^2$.  To this end it is convenient to introduce light-cone
vectors $n, \bar{n}$ which are defined through the four momenta of
virtual photon and target. With the Sudakov decomposition of a general
four vector $v$ given by
\begin{align}
  \label{eq:1}
  v_\mu & =  v^+ \bar{n}_\mu + v^- {n}_\mu + v_t\, , & \text{where} &&
    n\cdot \bar{n}&=1, & n^2& = 0 = \bar{n}^2 \, ,
 \notag \\
v^+ & = n \cdot v, \quad v^- = \bar{n} \cdot v \, , & \text{and} &&
 & v_t^2 = -{\bm v}^2\,,
\end{align}
we obtain for the momenta of  initial particles
\begin{align}
  \label{eq:mom-initial}
  P_\mu & = P^- n_\mu \, ,& l_\mu & = l^+ \bar{n} - \frac{Q^2}{l^+} n \, .
\end{align}
To include the possibility of arbitrary large gluon densities in the
target, we represent the latter by its gluonic field which can reach a
maximum strength of $A_\mu \sim 1/g$,  with $g$ the gauge coupling. To
calculate scattering amplitudes in the high energy limit it is then
convenient to treat the gluon field of the target as a background
field (shock-wave); in light-cone gauge $A \cdot n = 0$, the only
non-zero component is $A^-(x^+, x_t) = \delta(x^+) \alpha(x_t)$, while
$A_t = 0$  in the high energy limit.  Amplitudes are written in
terms of momentum space quark and gluon propagators in the presence of
the background field, see {\it e.g.} \cite{McLerran:1994vd},
\begin{align}
\label{eq:prop-bg}
S_{F, il} (p,q) &\equiv
 S_{F, il}^{(0)} (p) (2 \pi)^4 \delta^{(4)}(p - q) +
S_{F, ij}^{(0)}(p)  \, \cdot \,  \tau_{F, jk} (p,q) \, \cdot  \, S_{F, kl}^{(0)} (q)\, ,
\notag \\
G_{\mu\nu}^{ad} (p,q) &\equiv  G^{(0),ab}_{\;\;\mu\nu} (p) (2 \pi)^4 \delta^{(4)}(p - q) + G^{(0),ab}_{\;\;\mu\lambda} (p) \, \cdot \,  \tau_G^{bc} (p,q)  \,  \cdot \,   G^{(0), cd, \lambda}_{\;\;\;\; \nu} (q) \, ,
\end{align}
which are directly obtained from Fourier transforming their
corresponding counter parts in configuration space. In the above we
use the conventional free fermion and gluon propagator,
\begin{align}
  \label{eq:4}
  S_{F, ij}^{(0)} (p) & = \frac{i \delta_{ij}} {  (\slashed{p} + i\epsilon)}
&&\text{and}&
G^{(0), ab}_{\;\;\mu\nu}& =  \frac{i\delta^{ab}  d_{\mu\nu} (k)}
  {(k^2+i\epsilon)}
\end{align}
where
\begin{align}
  \label{eq:5}
d_{\mu\nu} (k)  &= - g_{\mu\nu} + \frac{k_\mu n_\nu + k_\nu n_\mu}{
    n\cdot k}
\end{align}
denotes the polarization tensor in the light-cone gauge, and
\begin{align}
  \label{eq:quarkinteraction}
\parbox{3cm}{\includegraphics[width=3cm]{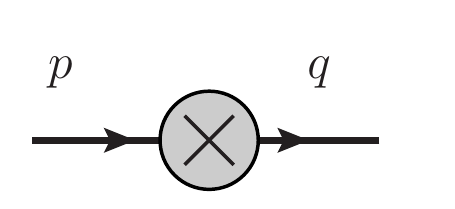}}
&  =  \tau_{F, ij}(p,q)   = 2 \pi \delta(p^+ - q^+) ~  \slashed{n}
\notag \\
&  \times \int d^{2} {\bm z} e^{i{\bm  z} \cdot ({ \bm p} - { \bm q})}
 \left\{\theta(p^+) \big[ V_{ij}({\bm z}) -1_{ij} \big]
-
\theta(-p^+) \big[ V_{ij}^\dagger({\bm z}) -1_{ij} \big]
  \right\} \\
\label{eq:gluoninteraction}
\parbox{3cm}{\includegraphics[width=3cm]{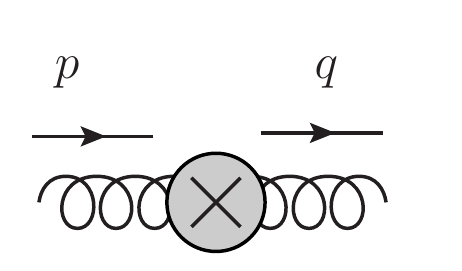}}
& =
 \tau_{G}^{ab}(p,q) =
2 \pi \delta(p^+ - q^+) ~ ( - 2 p^+)
\notag \\
&  \times
\int d^{2} {\bm z} e^{i{\bm z} \cdot ({\bm p} - {\bm q})}
 \left\{\theta(p^+) \big[ U^{ab}({\bm z}) -1 \big]
-
\theta(-p^+) \big[ \left(U^{ab}\right)^\dagger({\bm z}) -1 \big]
  \right\}
\end{align}
with Wilson lines in fundamental ($V$) and adjoint ($U$)
representation. They read
\begin{align}
  \label{eq:wilson}
  V(\bm z) & \equiv V_{ij} (\bm z) \equiv \mathrm{P} \exp ig \int\limits_{-\infty}^\infty
 d x^+ A^{-,c}(x^+, {\bm z})t^c \notag \\
 U(\bm z) & \equiv U^{ab} (\bm z) \equiv \mathrm{P} \exp ig \int\limits_{-\infty}^\infty
d x^+ A^{-,c}(x^+, {\bm z})T^c
\end{align}
with $-iT^{c}_{ab} = f^{acb}$.  To construct amplitudes in the
presence of a (strong) background field, it is  convenient to
extend conventional QCD momentum space Feynman rules by two additional
rules: (a) adding the vertices Eqs.~\eqref{eq:quarkinteraction} and
\eqref{eq:gluoninteraction} and (b) the requirement that all internal momenta $p$, {\it
  i.e.} momenta which cannot be expressed in terms of momenta of
external particles, are integrated over  with the measure $\int \frac{d^4 p}{(2 \pi)^4}$, in 1-1 correspondence to  conventional loop
momenta.  In tree diagrams such internal momenta arise if $n \geq 2$
vertices from Eqs.~\eqref{eq:quarkinteraction} and
\eqref{eq:gluoninteraction}, are inserted into a single Feynman
diagram; see Fig.~\ref{fig:example_mom_fdiag} for an illustrative
example.
If the number  $n$ of produced colored particles in the final state is
small, $n \leq 2$, the above method provides an efficient alternative
to the calculation of matrix elements in the presence of large gluon
densities, see \cite{JalilianMarian:2004da,Gelis:2002nn} for earlier
examples. For final states with large multiplicities, $n \geq 3$, the
method becomes inefficient due to the large number of Feynman diagrams
which need to be considered. While the process $\gamma^* +$ target
$\to q + \bar{q}$ requires 3 diagrams, one finds already 16 diagrams
for the process $\gamma^* +$ target $\to q + \bar{q} + g$.  Moreover,
calculations based on configuration space propagators suggest a large
redundancy in the result, leading to large cancellations among
different diagrams, which further complicate the use of momentum space
methods. In the following we demonstrate that there is a direct way to
use the emerging simplicity of the configuration space amplitudes for
calculations in momentum space.

\subsection{Cut diagrams according to negative and positive $x^+$}
\label{sec:cut-diagr-accord}

To achieve the required reduction of Feynman diagrams it is necessary
to exploit the restriction of the interaction between projectile
and target to the light-cone time-slice $x^+=0$,
$A^-(z) \sim \delta(z^+)$. Such properties are naturally revealed in
the configuration space and/or light-front formalism, which is the
main reason for their preferred use over momentum space
techniques. To make these properties and the associated reduction of
diagrams explicit in momentum space, it is sufficient to associate with
each propagator a separate minus-momentum variable. This can be
achieved by introducing at each (standard QCD) vertex (where
minus-momenta are conserved  {\it i.e.}, excluding the vertices in
Eqs.~\eqref{eq:quarkinteraction} and \eqref{eq:gluoninteraction}), a
delta function
$\delta(\{p^-_{\text{in}} \} - \{p^-_{\text{out}} \} )$, together with
a corresponding integration over the newly introduced minus momentum
variable; here $\{p_{\text{in (out)}} \}$ denotes the full set of
incoming (outgoing) momenta of the vertex. Expressing each of these
delta functions as a Fourier integral,
\begin{align}
  \label{eq:7}
  \delta(\{p^-_{\text{in}} \} - \{p^-_{\text{out}} \} ) & = \int \frac{d x^+}{2 \pi} e^{-i x^+ \cdot (\{p^-_{\text{in}} \} - \{p^-_{\text{out}} \}  )}\, ,
\end{align}
we finally obtain for each (standard QCD) vertex an integral over
$x_i^+$, $i = 1, \ldots N$ with $N$ the total number of (standard QCD)
vertices. These vertices are then connected by momentum space
propagators which are Fourier transformed w.r.t. their plus
momentum\footnote{The procedure is in 1-to-1 correspondence to the
  usual translation momentum space $\leftrightarrow$ configuration
  space, while in the current setup we limit ourselves to the minus
  momenta/plus coordinates}. At the vertices defined in
Eqs.~\eqref{eq:quarkinteraction}, \eqref{eq:gluoninteraction} the
minus momentum is not conserved and therefore no integration over
minus coordinates appears; instead these vertices are associated with
the `light-cone time' $x^+=0$, {\it i.e.} we use
$e^{i x^+ p^-}\bigg|_{x^+ = 0} = 1$. Fourier transforming quark and
gluon propagators w.r.t. to their plus momenta one finds easily the
well known expressions
\begin{align}
  \label{eq:6}
\tilde{S}_{F,kl}(x_{ij}^+; p^+, {\bm p}) & =  \int \frac{d p^-}{2 \pi} e^{- i p^- x_{ij}^+}S_{F,kl}(p)
=
\notag \\
 &   \hspace{-2.8cm} =   \delta_{kl} \frac{e^{- i p^- x_{ij}^+ }}{2 p^+} \bigg[ \bigg(
 \theta(p^+)\theta(x_{ij}^+)
+
 \theta(-p^+)\theta(- x_{ij}^+)\bigg) \left(\slashed{p} + m\right)
+ \delta(x_{ij}^+) \slashed{n}
  \bigg]_{p^- = \frac{{\bm p}^2 + m^2}{2 p^+}} \notag  \\
\tilde{G}^{(0),ab}_{\mu\nu}(x_{ij}^+; p^+, {\bm p}) & =   \int \frac{d p^-}{2 \pi} e^{- i p^- x_{ij}^+}G_{\mu\nu}^{(0),ab}(p) =
\notag \\
 & \hspace{-2.8cm}= \delta_{ab} \frac{e^{- i p^- x_{ij}^+ }}{2 p^+} \bigg[ \bigg(
 \theta(p^+)\theta(x_{ij}^+)
+
 \theta(-p^+)\theta(- x_{ij}^+)\bigg)  \cdot  d_{\mu\nu}(p)
+  2 \delta(x_{ij}^+) \frac{n_\mu n_\nu}{p\cdot n}
  \bigg]_{p^- = \frac{{\bm p}^2 + m^2}{2 p^+}} \, .
\end{align}
Here $x_{ij}^+ \equiv x_i^+ - x_j^+$.  Using such propagators together
with integrations over light-cone times $x^+_i$ at each vertex one
re-obtains directly the light-cone time ordered Feynman rules of
light-front perturbation theory (old-fashioned perturbation theory),
see {\it e.g.} \cite{Beuf:2016wdz} which contains a recent and compact
overview. In particular we find that it is natural to organize QCD
interaction vertices into vertices before ($x_i^+<0$) and after the
interaction ($x_i^+ > 0$) with the target field.  In a nut-shell this
result will allow us to evaluate to zero configurations which are
absent in the light-front formalism for momentum space diagrams,
simplifying enormously the reduction of diagrams at earlier stages of
the calculation.\\

\begin{figure}[t] 
\centering
 \parbox{.4\textwidth}{\center \includegraphics[width
=.4\textwidth]{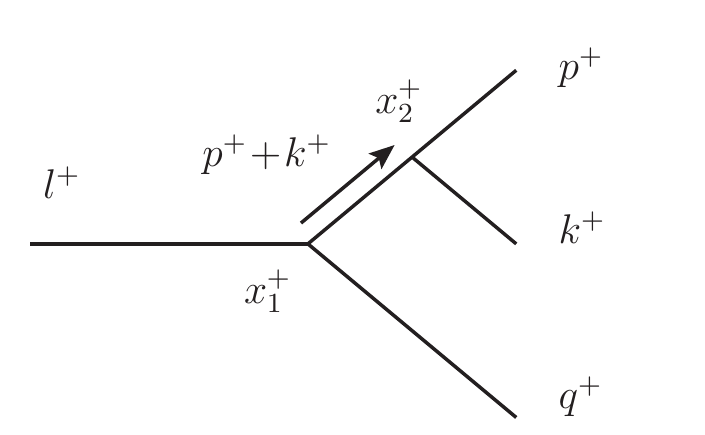}}
\parbox{.4\textwidth}{\center \includegraphics[width
=.4\textwidth]{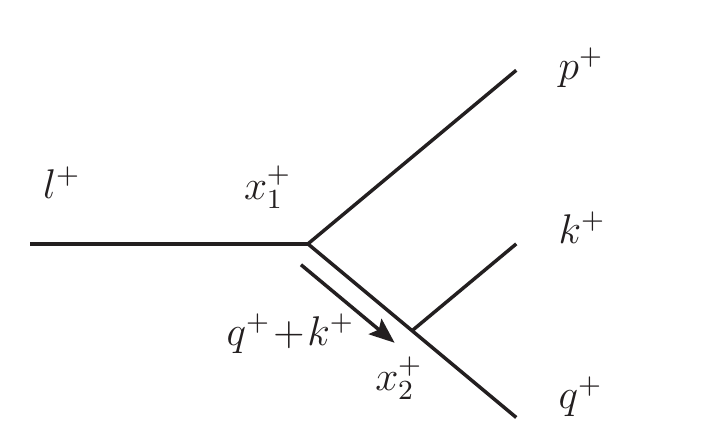}}

\parbox{.4\textwidth}{\center (a)}\parbox{.4\textwidth}{\center (b)}

  \caption{\it Real corrections. We have $l^+, p^+, q^+, k^+ >0$ and
$l^+ = p^+ + k^+ + q^+$}
  \label{fig:real}
\end{figure}

\begin{figure}[h] \centering
\parbox{.3\textwidth}{\center \includegraphics[width
=.3\textwidth]{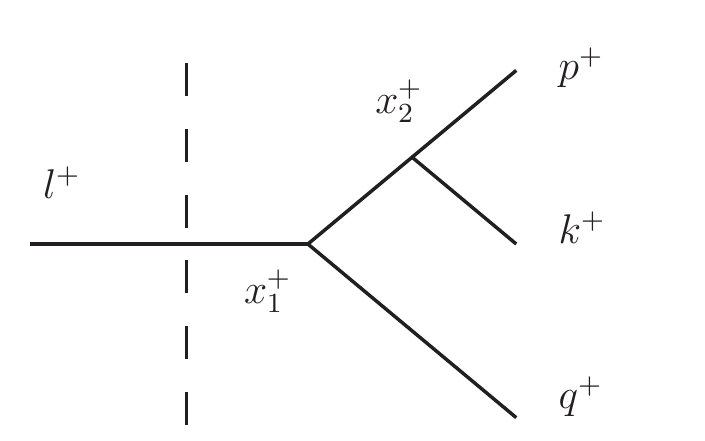}}
\parbox{.3\textwidth}{\center \includegraphics[width
=.3\textwidth]{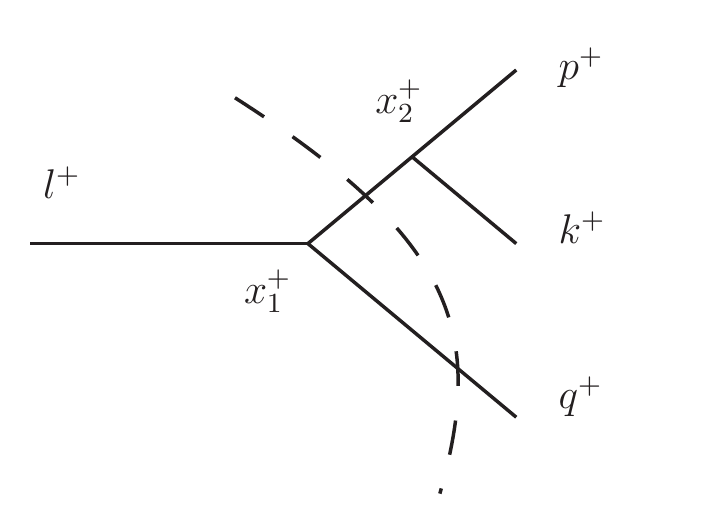}}
\parbox{.3\textwidth}{\center \includegraphics[width
=.3\textwidth]{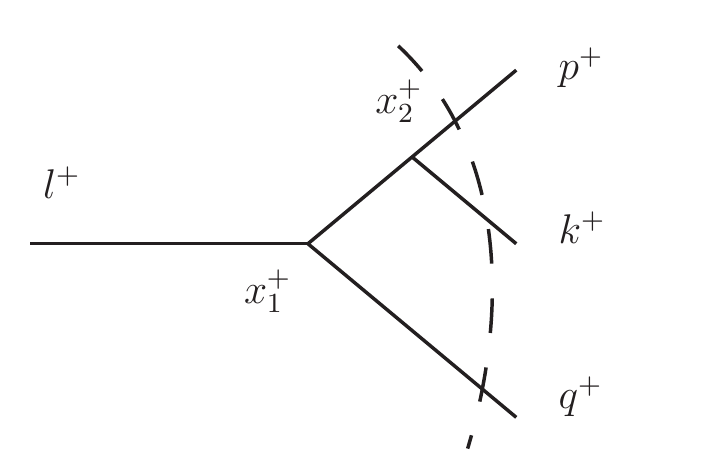}}
\\
\parbox{.3\textwidth}{\center \includegraphics[width
=.3\textwidth]{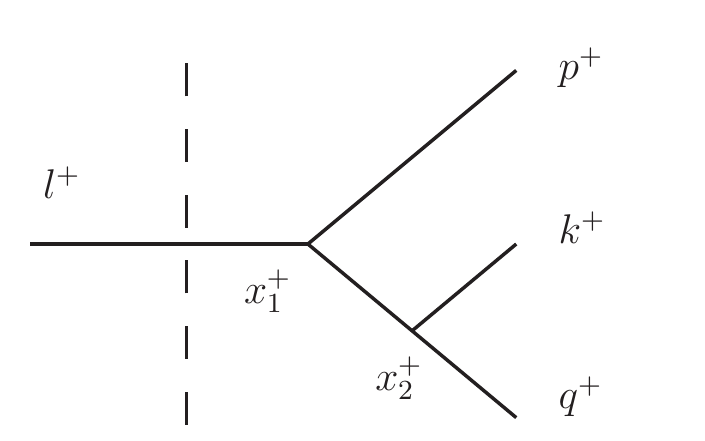}}
\parbox{.3\textwidth}{\center \includegraphics[width
=.3\textwidth]{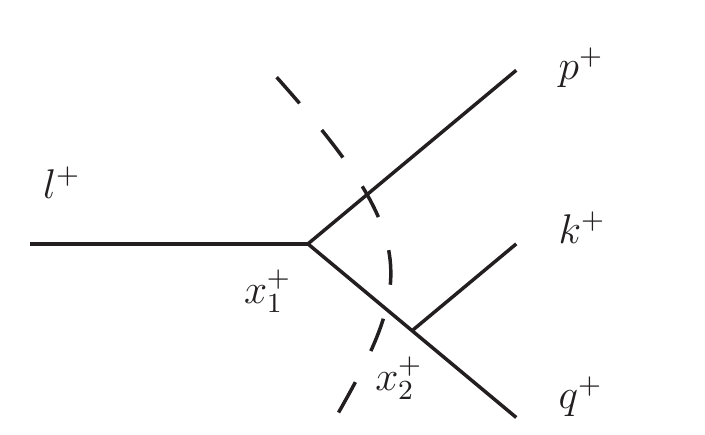}}
\parbox{.3\textwidth}{\center \includegraphics[width
=.3\textwidth]{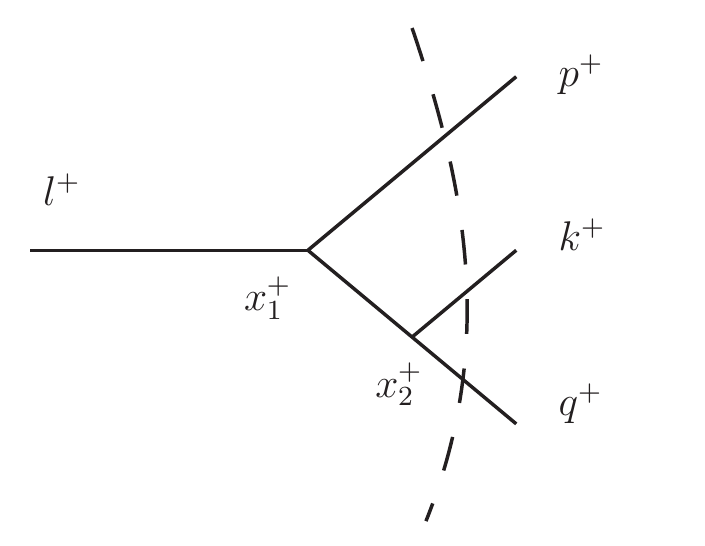}}
  \caption{\it Real corrections. The dashed line indicates the $x^+=0$
time-slice where the interaction with the target can take place. In
addition there are also non-interacting contributions}
    \label{fig:real_sliced}
\end{figure}

To be more precise, we first note that potential contact terms in the
quark ( $\sim \delta(x_{ij}^+) \slashed{n}$) and gluon
($\sim \delta(x_{ij}^+) n_\mu n_\nu$) propagators are absent, if the
propagators connect with the vertices 
Eqs.~\eqref{eq:quarkinteraction} and \eqref{eq:gluoninteraction}: for
the quark due to the identity $\slashed{n} \slashed{n} = 0$, for the
gluon since $d_{\mu\nu}(p) \cdot n^\nu = 0$ and $n^2 = 0$ (if the
vertex connects to a virtual gluon) and
$\epsilon^{(\lambda)}(p)_\nu\cdot n^\nu = 0$ (if the vertex connects
to a real gluon).  Due to the absence of such contact terms it will be
sufficient to limit the discussion to scalar skeleton diagrams,
independent of the precise nature of particles. We further note that
plus momenta are conserved at all vertices (including
Eqs.~\eqref{eq:quarkinteraction} and \eqref{eq:gluoninteraction}). On
the other hand, since plus coordinates and plus momenta are the only
components which receive special constraints from the propagators in
the representation of Eq.~\eqref{eq:6}, it is sufficient to study only
their effect. The resulting diagrams are depicted in
Fig.~\ref{fig:real}, where for the time being we ignore the presence
of potential target interaction vertices
Eqs.~\eqref{eq:quarkinteraction} and \eqref{eq:gluoninteraction}.
In a next step we divide all integrals over light-cone times $x_i^+$
into a positive $x_i^+>0$ and a negative $x_i^+ < 0$ sector. While
this is motivated by the restriction of the target interaction vertices
to light-cone time $x_i^+ = 0$, this is irrelevant for the following
reason. Using now that propagation from a vertex with negative
light-cone time $x_j^+ < 0$ to a positive light-cone time $x_i^+ > 0$
implies $x_{ij}^+ > 0$ and therefore positive light-cone momentum
$p^+>0$ (due to the structure of theta-functions in
Eq.~\eqref{eq:6}), it is straight forward to verify that each skeleton
diagram can be organized according to three possible ``$s$-channel'' cuts, see
Fig.~\ref{fig:real_sliced}.
It is easily possible to extend our result to the case of $n$ final
state particles: crossing from negative to positive light-cone times
requires positive light-cone momentum and -- vice versa -- crossing
from positive to negative light-cone times requires negative
light-cone momentum. In the case of a tree skeleton diagram, all
light-cone momenta are always positive and therefore crossing from
positive to negative light-cone times is impossible. Moreover, once a
certain line crosses from the negative to the positive light-cone time
sector, all its daughter lines, {\it i.e.}, lines emerging from a
splitting of this line, will have positive light-cone times by
default.  The resulting picture tells us that each individual diagram
is characterized by all possible vertical lines through the diagrams
(`$s$-channel' cuts)
which indicate transition from negative to positive light-cone time. \\

\begin{figure}[t]
  \centering
  \includegraphics[width=.25\textwidth]{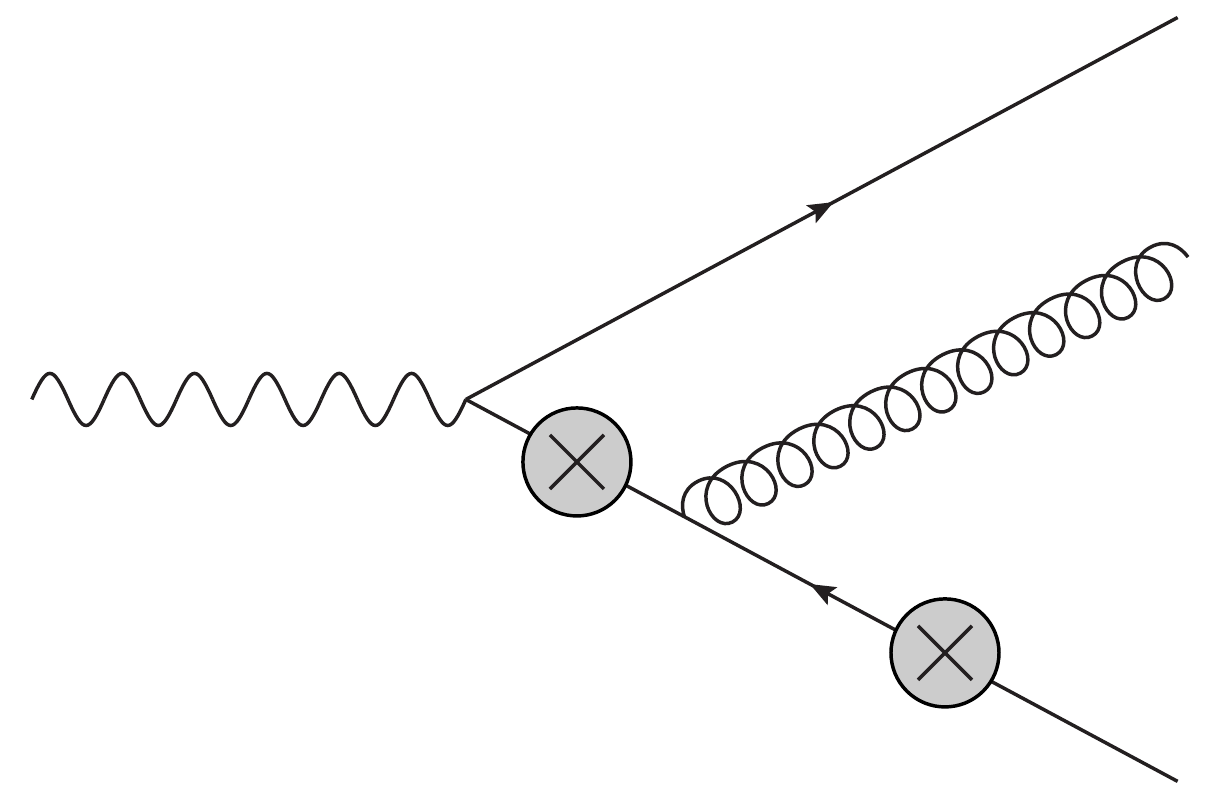}
  \hspace{1cm} 
  \includegraphics[width=.25\textwidth]{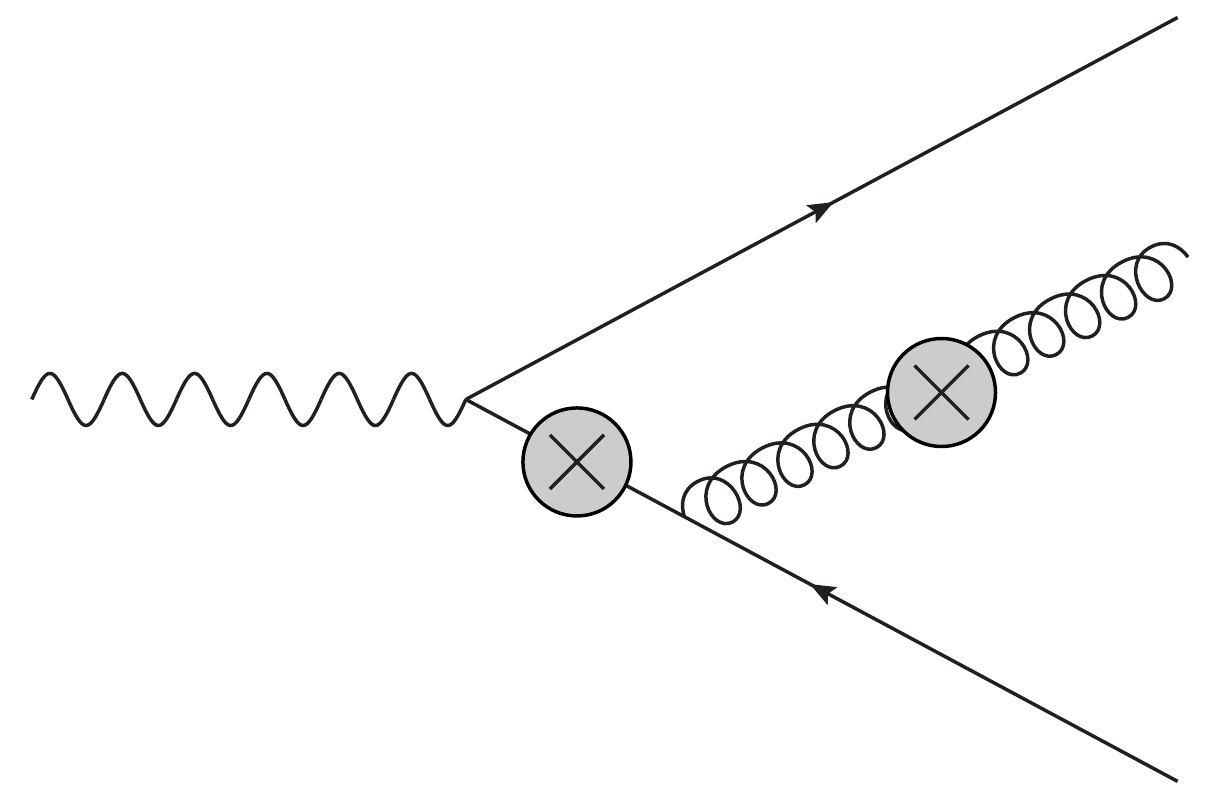}
  \caption{Diagrams with interaction not aligned along a vertical cut and which gives therefore a zero contribution}
  \label{fig:illegal}
\end{figure}

\begin{figure}[t]
  \centering
\includegraphics[width=.25\textwidth]{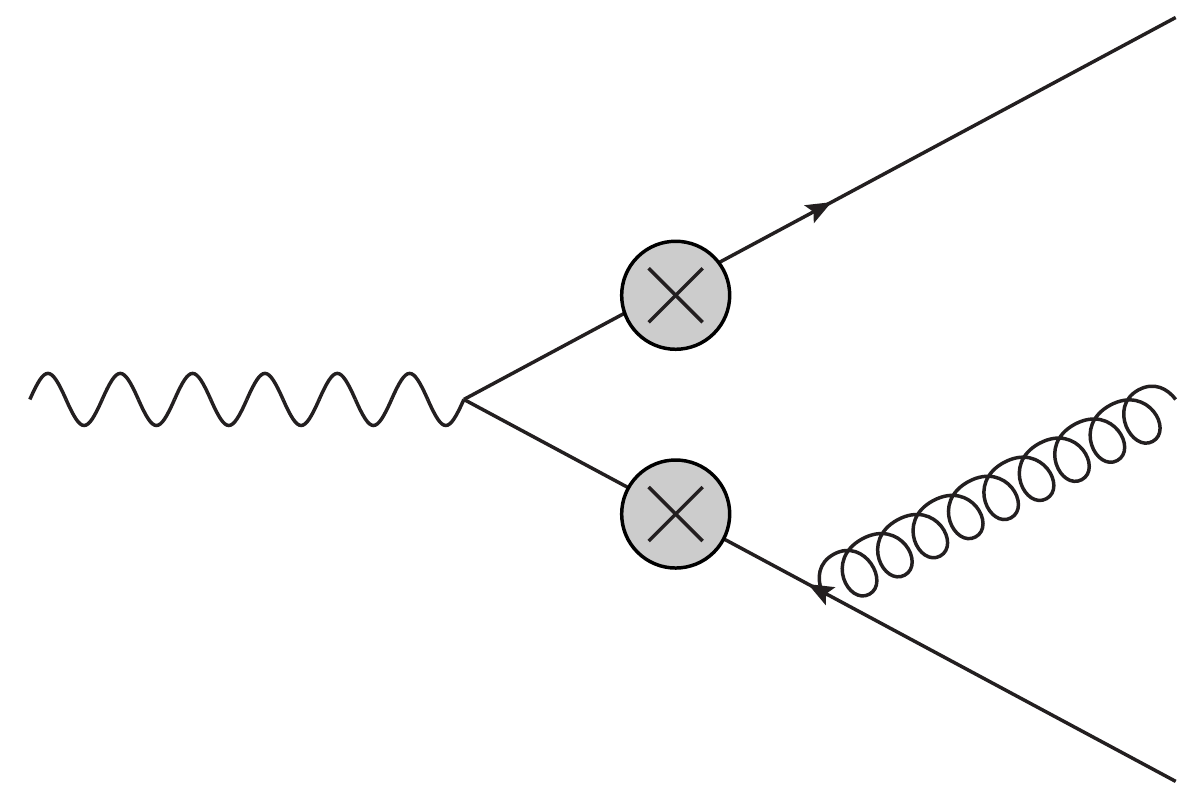} \hspace{1cm}
\includegraphics[width=.25\textwidth]{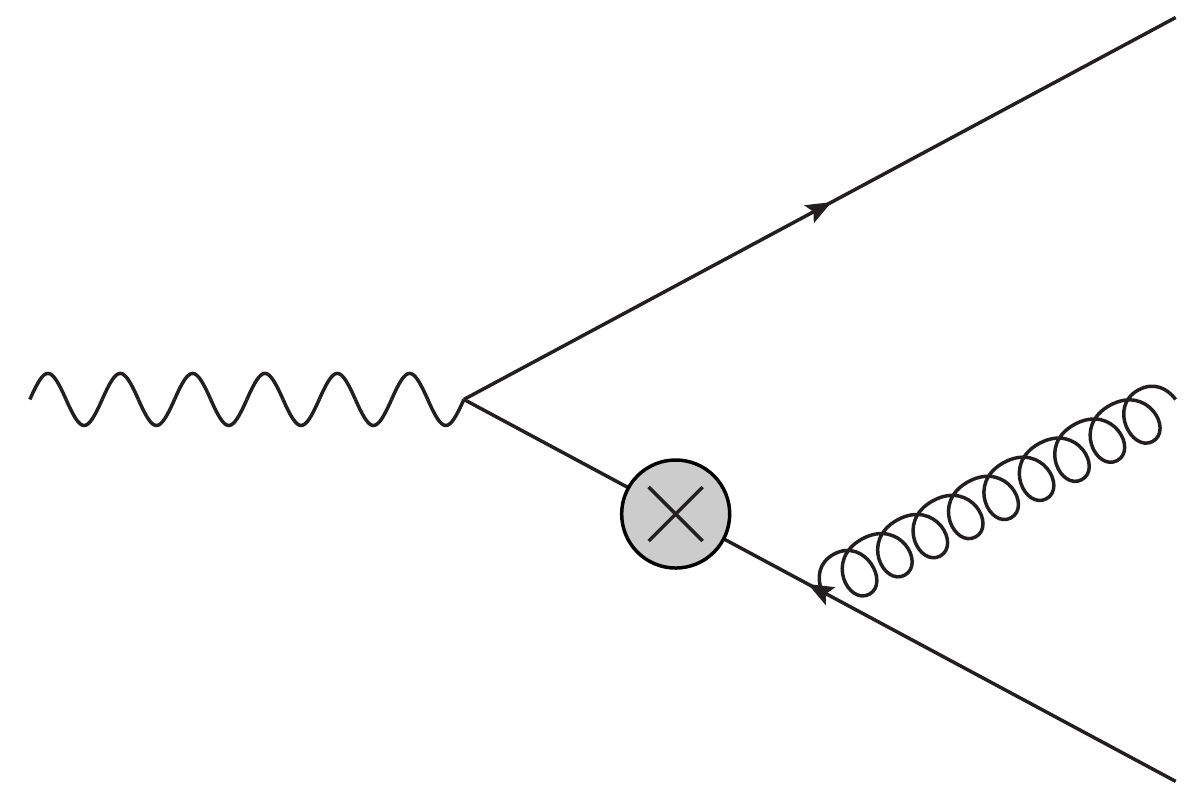} \hspace{1cm}
\includegraphics[width=.25\textwidth]{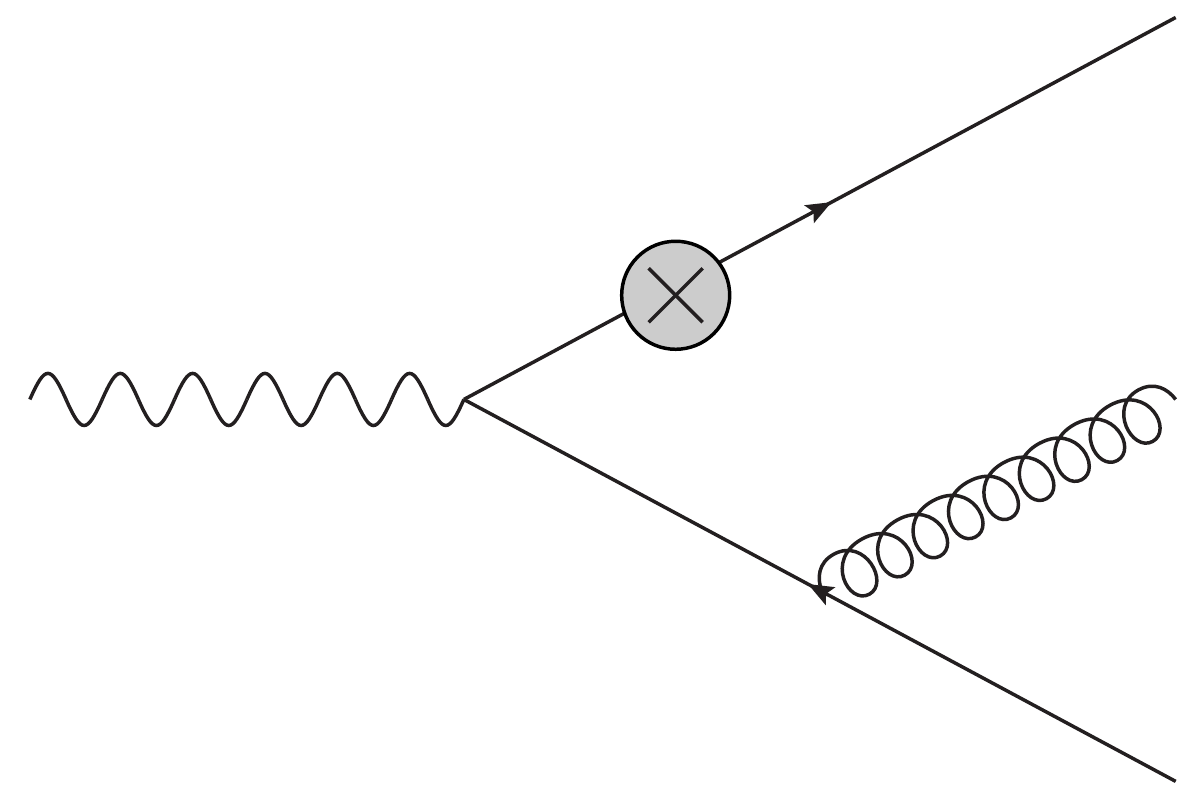}
 
  \caption{Diagrams with and without interaction which belong to the same cut. }
 \label{fig:cut_ex}
\end{figure}

The benefit of this result for the study of the interaction with the
target field should be apparent by now: since vertices,
Eqs.~\eqref{eq:quarkinteraction} and \eqref{eq:gluoninteraction}, are
limited to light-cone time $x_i^+=0$, such vertices can only be
inserted in a ``cut'' line. Insertions in un-cut lines, are
immediately zero, see Fig.~\ref{fig:illegal} for two configurations
which cannot occur.
Note that this result applies separately for each individual Feynman
diagram and holds regardless of whether the actual evaluation takes
place in configuration, momentum or mixed {\it i.e.} light-front
space.  While this leads already to a significant reduction in the
number of Feynman diagrams to be evaluated, the remaining set of
diagrams still contains a sizable fraction of redundant
contributions. In particular there are large cancellations between
diagrams where we place a target interaction vertex
Eqs.~\eqref{eq:quarkinteraction}, \eqref{eq:gluoninteraction} at the
$z_i^+=0$ cut and diagrams where such interaction is absent, see
Fig.~\ref{fig:cut_ex} for an example.\\

\subsection{Further reduction of diagrams}
\label{sec:further-reduction}

To reduce the number of diagrams further, it is necessary to study the
Fourier transform of the complete propagators (containing both
interacting and non-interacting) parts, Eq.~\eqref{eq:prop-bg}. Using these
propagators for all lines in diagrams (for external lines this implies
the use of the LSZ-reduction procedure, see {\it e.g.}
\cite{JalilianMarian:2004da}) the full process can be represented in
terms of two diagrams only.  On the level of scalar skeleton graphs
these are precisely those of Fig.~\ref{fig:real}. It is well known
that such complete propagators can be written in configuration space
in three parts: one term is associated with crossing from
negative to positive light-cone time $x^+$ and is directly
proportional to either a Wilson line (positive light-cone momentum
fraction) or a hermitian conjugate Wilson line (negative light-cone
momentum fraction). The other two terms describe free propagation
between two points with either negative or positive light-cone
time. For the following discussion the following form is sufficient:
\begin{align}
  \label{eq:complete_prop}
 &  \int \frac{d p^-}{2 \pi}  \int \frac{d q^-}{2 \pi} e^{- i p^- x^+} e^{i q^- y^+} \bigg[ 
 S_{F, il}^{(0)} (p) (2 \pi)^4 \delta^{(4)}(p - q) +
S_{F, ij}^{(0)}(p)  \cdot   \tau_{F, jk} (p,q)  \cdot   S_{kl}^{(0)} (q)
\bigg] \notag \\
&=
(2 \pi)^3\delta(p^+ - q^+)   \delta^{(2)}(p_t - q_t) 
\tilde{S}^{(0)}_{F, il}(x^+ - y^+; p^+, p_t) \theta(x^+ \cdot  y^+) 
\notag \\
& \hspace{8cm}
+ \tilde{S}_{F, il}^{(V, V^\dagger)}(x^+, y^+; p^+, p_t ; q^+, q_t) \notag \\
 &  \int \frac{d p^-}{2 \pi}  \int \frac{d q^-}{2 \pi} e^{- i p^- x^+} e^{i q^- y^+} \bigg[ 
 G_{\mu\nu}^{(0),ab} (p) (2 \pi)^4 \delta^{(4)}(p - q) +
G_{\mu\alpha}^{(0),ac}(p)  \cdot   \tau_{G}^{cd} (p,q)  \cdot   G_{\alpha\nu}^{(0),db} (q)
\bigg] \notag \\
&=
(2 \pi)^3\delta(p^+ - q^+)   \delta^{(2)}(p_t - q_t) 
\tilde{G}_{\mu\nu}^{(0),ab}(x^+ - y^+; p^+, p_t) \theta(x^+ \cdot  y^+) 
\notag \\
& \hspace{8cm}
+ \tilde{G}_{\mu\nu}^{ab, (U, U^\dagger)}(x^+, y^+; p^+, p_t ; q^+, q_t)
\end{align}
where $\tilde{S}^{(0)}_{F,il}$, $\tilde{G}^{(0,ab)}_{\mu\nu}$ are
given by Eq.~\eqref{eq:6}. The second term 
necessarily involves crossing the slice $x^+ = 0$ and is given by
\begin{align}
  \label{eq:SV_SVdagger}
 &  \tilde{S}_{F, il}^{(V, V^\dagger)}(x^+, y^+;  p^+, p_t ; q^+, q_t)  \notag \\ & = \theta(p^+) \tilde{S}_{F,ij}(x^+, p^+, p_t)  \slashed{n} \tilde{S}_{F,kl}(-y^+; p^+, q_t)
 \cdot 2 \pi \delta (p^+ - q^+)\int d^2 z_t e^{i z_t \cdot (p_t - q_t)} V_{jk}(z_t) 
\notag \\
& - 
 \theta(-p^+) \tilde{S}_{F,ij}(x^+, p^+, p_t)  \slashed{n} \tilde{S}_{F,kl}(-y^+; p^+, q_t)
 \cdot 2 \pi \delta (p^+ - q^+)\int d^2 z_t e^{i z_t \cdot (p_t - q_t)} V^\dagger_{jk}(z_t) 
\notag \\
& \hspace{6cm} =  \tilde{S}_{F,ij}(x^+, p^+, p_t) \cdot \bar{\tau}_{F,jk}(p,q) \cdot   \tilde{S}_{F,kl}(-y^+; p^+, q_t) 
\end{align}
where the condition $p^+ > 0 $ ($p^+<0$) selects automatically the configuration $x^+ > 0 > y^+$ ($x^+ < 0 < y^+$). Similarly one has in the case of gluons
\begin{align}
  \label{eq:G_UUdagger}
 &  \tilde{G}_{\mu\nu}^{ab, (U, U^\dagger)}(x^+, y^+;  p^+, p_t ; q^+, q_t) 
 = 
 \tilde{G}_{\mu\alpha}^{(0), ac}(x^+, p^+, p_t)  \bar{\tau}_G^{cd}(p,q) 
\tilde{G}_{\alpha\nu}^{(0), db}(-y^+; p^+, q_t)
\end{align}
where we defined 
\begin{align}
  \label{eq:quarkinteraction2}
\parbox{3cm}{\includegraphics[width=3cm]{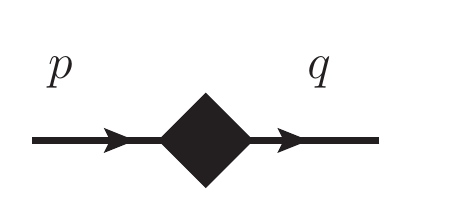}}
&  =  \bar{\tau}_{F, ij}(p,q)   = 2 \pi \delta(p^+ - q^+) \cdot  \slashed{n}
\notag \\
&  \cdot \int d^{2} {\bm z} e^{i{\bm z} \cdot ({\bm p} - {\bm q})}
 \left\{\theta(p^+)  V_{ij}({\bm z})
-
\theta(-p^+)  V_{ij}^\dagger({\bm z})
  \right\} \\
\label{eq:gluoninteraction2}
\parbox{3cm}{\includegraphics[width=3cm]{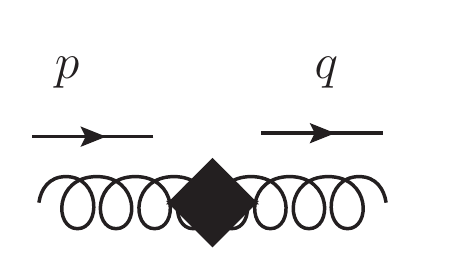}}
& =
 \bar{\tau}_{G}^{ab}(p,q) =
2 \pi \delta(p^+ - q^+) \cdot ( - 2 p^+)
\notag \\
&  \cdot
\int d^{2} {\bm z} e^{i{\bm z} \cdot ({\bm p} - {\bm q})}
 \left\{\theta(p^+)  U^{ab}({\bm z})
-
\theta(-p^+)  \left(U^{ab}\right)^\dagger({\bm z})
  \right\}
\end{align}
Applying now these results to the (skeleton) diagrams,
Fig.~\ref{fig:real_sliced}, we now find that each ``cut'' line must
necessarily come with a vertex, Eqs.~\eqref{eq:quarkinteraction2}
and~\eqref{eq:gluoninteraction2}. The ``un-cut'' lines can only come
with a free propagator. At first these propagators are limited to
positive or negative light-cone time only and therefore cannot be directly related to their momentum space counter-parts. Due to the results of
Sec.~\ref{sec:cut-diagr-accord}, adding a free propagator which
crosses from negative to positive light-cone time will however give
only a zero contribution. Adding such a zero contribution it is then
straight forward to Fourier transform our result back to momentum
space. Therefore the complete amplitude can be calculated from the six
diagrams of Fig.~\ref{fig:real_sliced} with a vertex,
Eqs.~\eqref{eq:quarkinteraction2} and~\eqref{eq:gluoninteraction2} at
each cut, minus the diagram with the background field $A^+$ set to
zero; if the initial particle is is not colored (as in our case) the
total number of diagrams reduces finally to four.

\subsection{The minimal set of amplitudes}
\label{sec:minim-set-ampl}
\begin{figure}[t]
  \begin{center}
    \includegraphics[width=.38\textwidth]{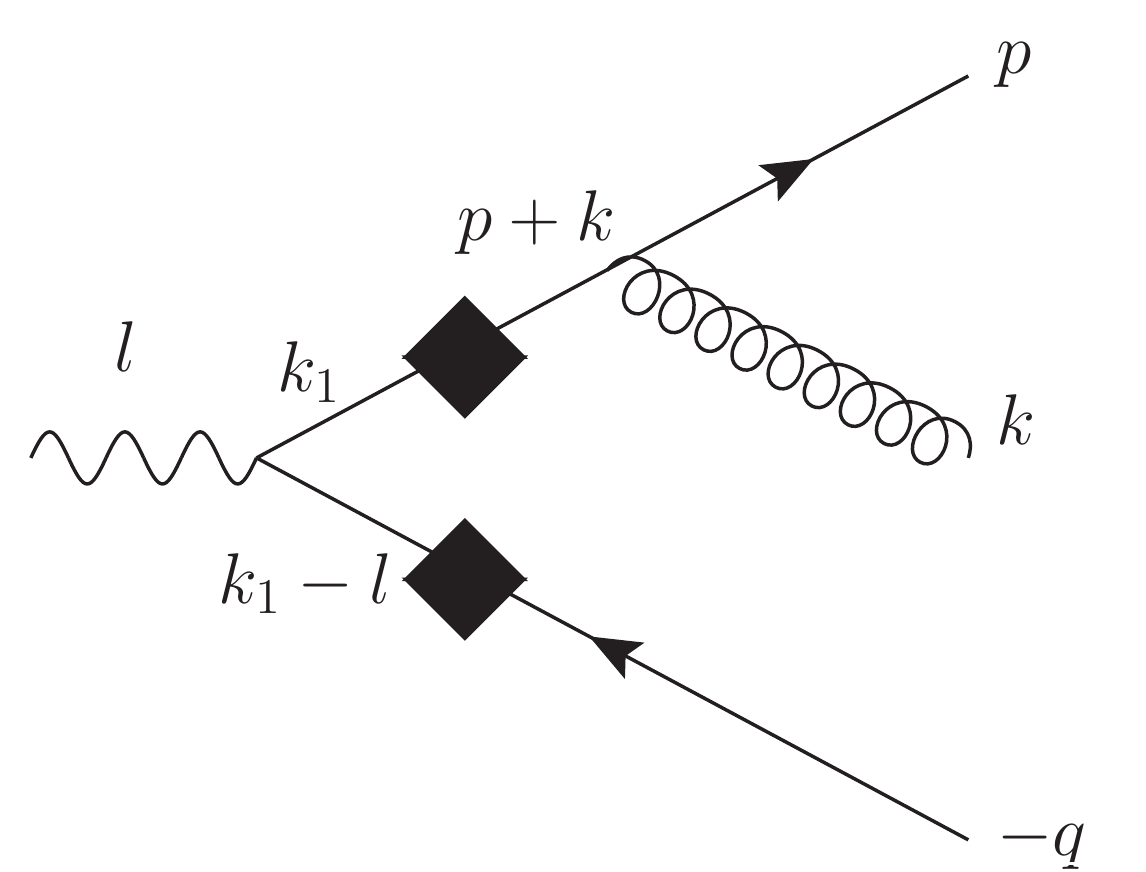}
\includegraphics[width=.38\textwidth]{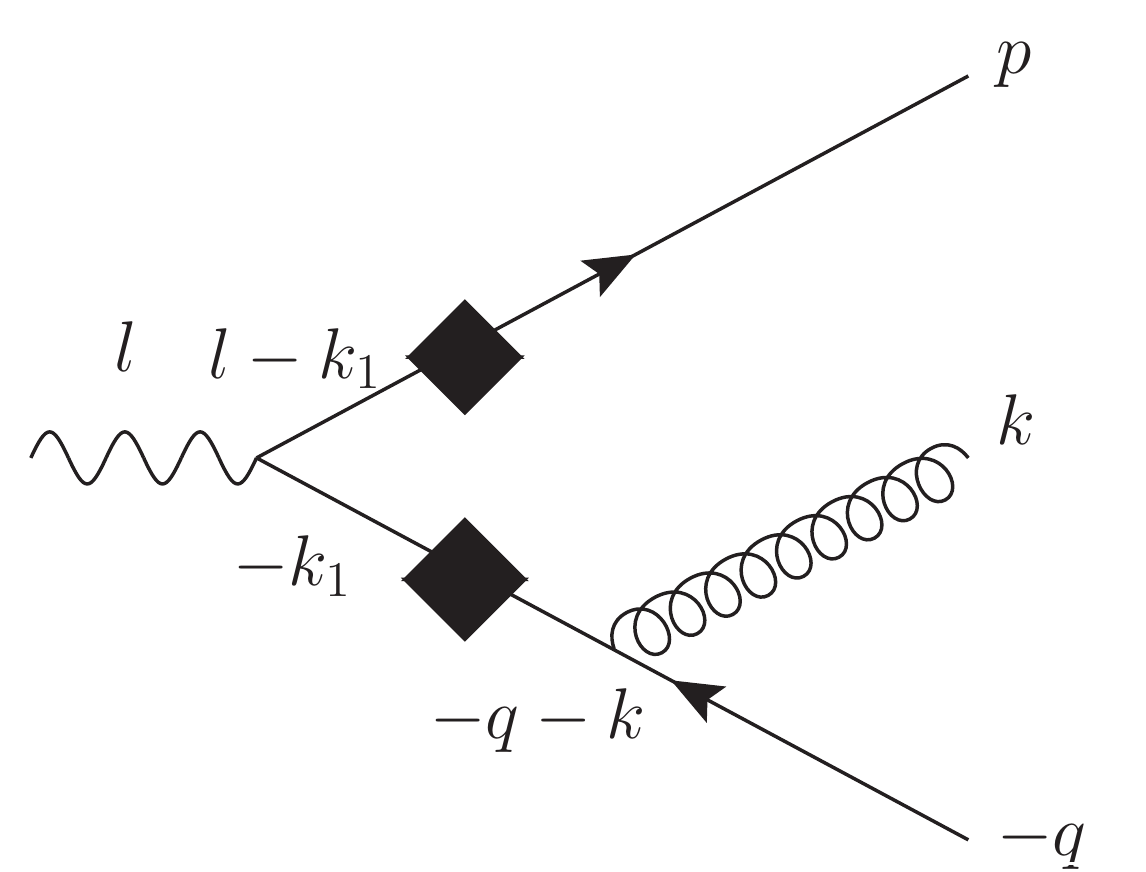}\\
\includegraphics[width=.38\textwidth]{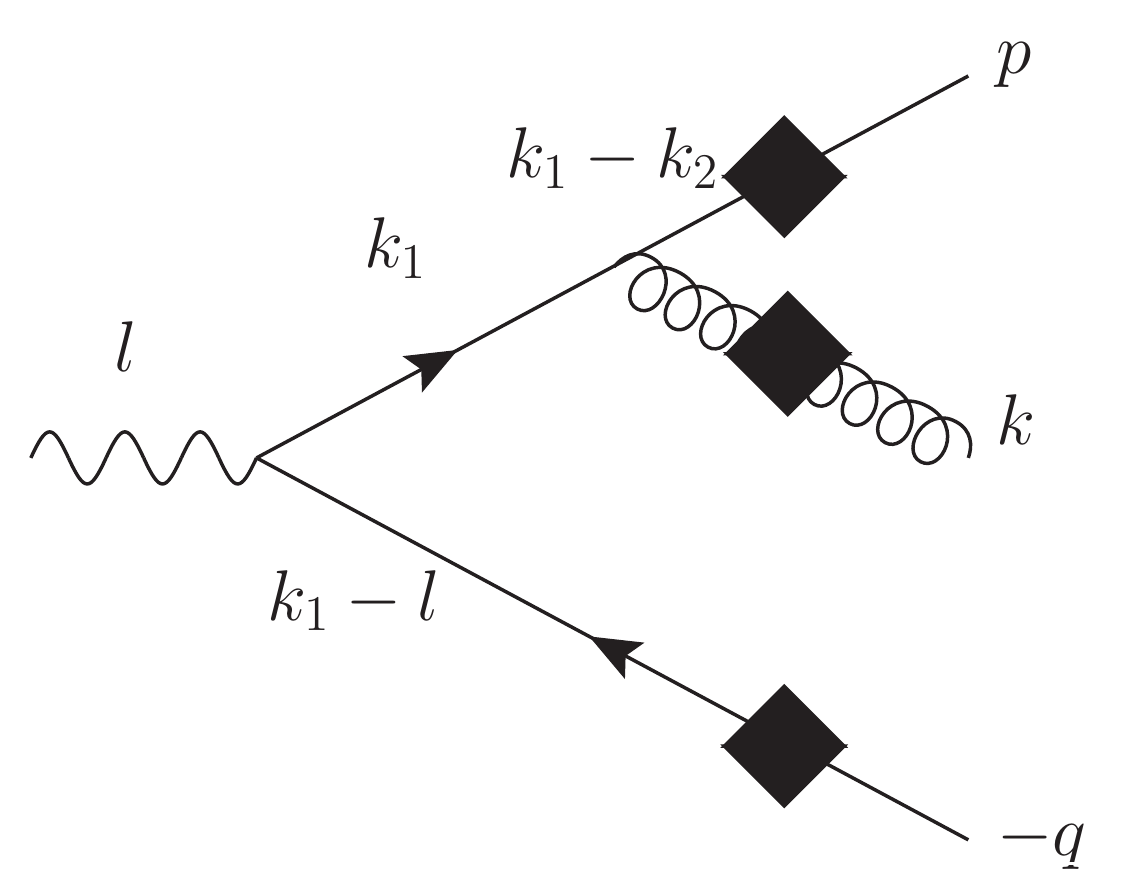}
\includegraphics[width=.38\textwidth]{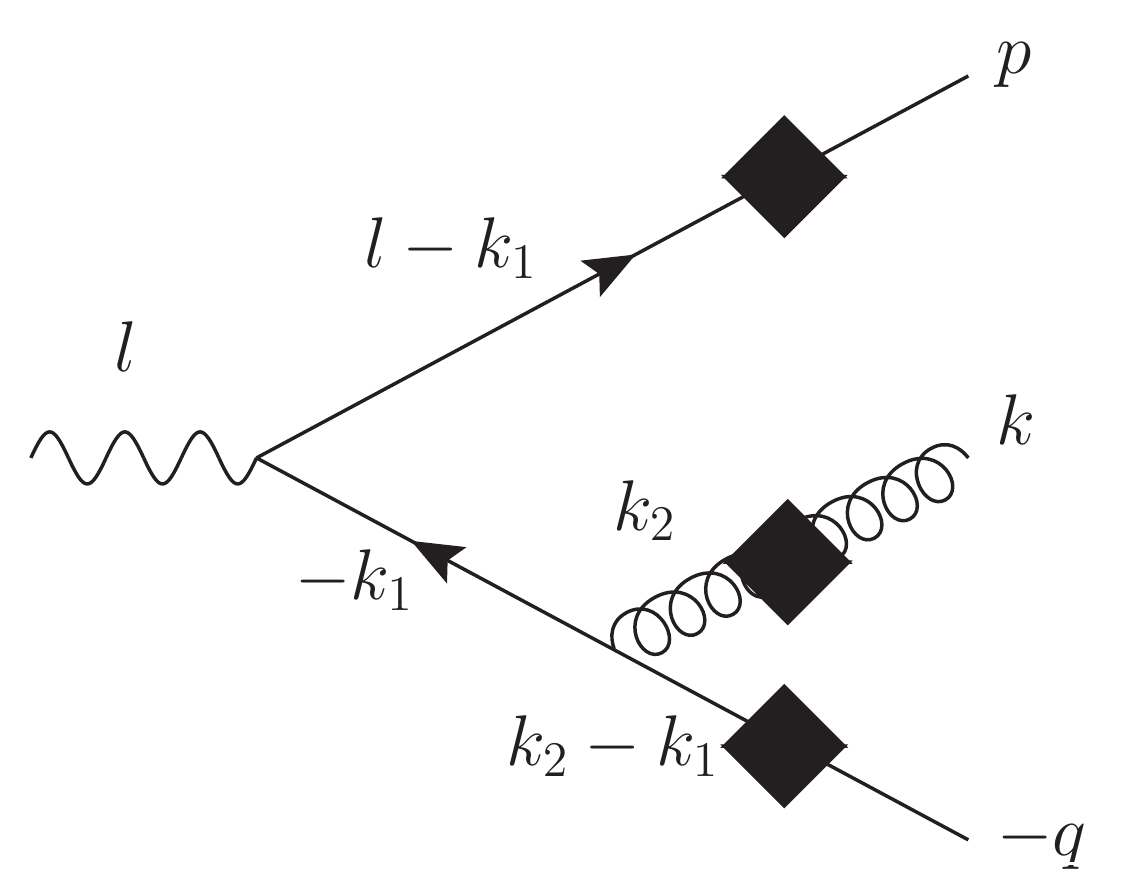}
  \end{center}

\caption{\small  $3$-parton production diagrams.   The arrows indicate the direction of fermion charge flow. The photon momentum is incoming whereas all the final state momenta are outgoing.
}
\label{fig1}
\end{figure}
The final set of diagrams which we need to evaluate for the process
described by Eq.~\eqref{eq:process} is depicted in Fig.~\ref{fig1}.
The four matrix elements corresponding to Fig.~\ref{fig1} read
\begin{align}
  i{\cal A}_1 = &  (i e) (i g)  \int {d^4 k_1 \over (2\pi)^4} \bar{u}(p)   \gamma^\mu\, t^a\,
\overline{S}_F (p+k, k_1)  \gamma^\nu
  \overline{S}_F (k_1-l, -q)
 \notag \\
& \hspace{6cm} \cdot  \big[S_F^{(0)} (-q)\big]^{-1}\, v(q) \epsilon_\nu (l) \,
   \epsilon^*_\mu (k)\, ,
 \notag \\
i{\cal A}_2 =&  (i e) (i g) \int  {d^4 k_1 \over (2\pi)^4} \bar{u} (p) \, \big[S_F^{(0)} (p)\big]^{-1}\overline{S}_F (p, k_1)\,
  \gamma^\nu \overline{S}_F (k_1-l,-q-k) \, 
\notag \\
& \hspace{6cm} \cdot 
\gamma^\mu\, t^a\,  v(q) \,,
  \epsilon_\nu (l) \, \epsilon^*_\mu (k)\,, \notag \\
i{\cal A}_{3}   =&   (i e) (i g) \int {d^4 k_1 \over (2\pi)^4} \int
                 {d^4 k_2 \over (2\pi)^4} \bar{u} (p) \big[S_F^{(0)}
                 (p)\big]^{-1}
  \overline{S}_F (p, k_1-k_2)
\gamma^\lambda\,t^c\, S_F^{(0)} (k_1)
 \gamma^\nu
\notag \\
&   \overline{S}_F (k_1-l,-q)  \big[S_F^{(0)} (-q)\big]^{-1}\!\!  v(q)
 \left[\overline{G}_{\lambda}^{\;\,\delta}\right]^{ca}\!\! (k_2,k)
 \big[G_{\delta}^{(0), \mu} (k) \big]^{-1}   \epsilon_\nu (l)  \epsilon_\mu^* (k)\,  , \notag \\
i{\cal A}_{4} =&  (i e) (i g)
 \int {d^4 k_1 \over (2\pi)^4} \int {d^4 k_2 \over (2\pi)^4}
 \bar{u} (p) \big[S_F^{(0)} (p)\big]^{-1} \overline{S}_F (p, l-k_1)\,
\gamma^\nu\, S_F^{(0)} (-k_1) \gamma^\lambda\, t^c
\notag \\
&
\overline{S}_F (k_2-k_1,-q)
 \big[S_F^{(0)} (-q)\big]^{-1}\! \left[\overline{G}^{\;\,\delta}_\lambda\right]^{ca}\! (k_2,k)
  \big[G^{(0) \mu}_\delta (k) \big]^{-1}
\epsilon_\nu (l)  \,
 \epsilon^*_\mu (k),
\label{eq:calA1234}
\end{align}
where $\epsilon_\nu (l), \epsilon^*_\mu (k)$ denote polarization
vectors of the incoming virtual photon and the outgoing gluon
respectively and 
\begin{align}
\label{eq:8}
\overline{S}_{F, il} (p,q) &\equiv
S_{F, ij}^{(0)}(p)  \, \cdot \,  \bar{\tau}_{F, jk} (p,q) \, \cdot  \, S_{kl}^{(0)} (q),
\notag \\
\overline{G}_{\mu\nu}^{ad} (p,q) &\equiv 
                        G^{(0),ab}_{\;\;\mu\lambda} (p) \, \cdot \,  \bar{\tau}_G^{bc} (p,q)  \,  \cdot \,   G^{(0), cd, \lambda}_{\;\;\;\; \nu} (q)
\end{align}
With flux factor $\mathcal{F} = 2 l^+$, photon momentum fractions
$\{z_1, z_2, z_3 \} = \{p^+/l^+, q^+/l^+, k^+/l^+\}$, and the
three-particle phase space,
\begin{align}
  \label{eq:10}
   d\Phi^{(3)}& =\frac{1}{(2 \pi)^8} \,  d^4p  \, d^4 q  \,  d^4 k  \,
                  \delta(p^2)\delta(q^2)\delta(k^2)
\delta(l^+  -  p^+  -
                  k^+ - q^+) \notag \\
&=
2l^+  \frac{ d^2 {\bm p} \,  d^2 {\bm q}\,  d^2 {\bm k} }{(64 \pi^4 l^+)^2}
  \frac{dz_1  \,   dz_2 \,  dz_3}{z_1 \,
  z_2 \,  z_3 } \delta(1-\sum_{i=1}^3z_i)\, ,
\end{align}
 the differential 3-parton production  cross-section reads
\begin{align}
  \label{eq:2}
  d \sigma & = \frac{1}{\mathcal{F}} \left\langle   | \sum_{i=1}^4 \mathcal{M}_i(A^+) -  \mathcal{M}_i(0)|^2  \right\rangle_{A^+} d \Phi^{(3)},
\end{align}
where $\langle \ldots \rangle_{A_+}$ denotes the average over
background field configurations, $\mathcal{F} = 2 l^+$ and
\begin{align}
  \label{eq:24}
  \mathcal{A}_i  & = 2 \pi \delta (l^+ - p^+ - k^+ - q^+) \mathcal{M}_i & 
i&= 1, \ldots, 4\, .
\end{align}

\section{Spinor helicity methods}
\label{sec:helic-spin-meth}
While spinor helicity methods are well established for calculations
within conventional perturbative QCD calculations, its application to
calculations in the QCD high energy/Regge limit are rather limited,
see~\cite{anna} for some examples. We therefore start this section by
recalling basic definitions which will further serve to fix our
notation.

\subsection{Basic definitions}
\label{sec:basics}
The presentation in this paragraph follows closely those of the
reviews \cite{dixon}.  For massless fermions, helicity is a good, {\it
  i.e.} conserved, quantum number. One defines (on-shell) helicity
eigenstates (spinors) as
\begin{align}
  \label{eq:spinor_def}
  u_\pm(k) & = \frac{1 \pm \gamma_5}{2} u(p) & 
 v_\mp(k) & =  \frac{1 \pm \gamma_5}{2} v(p) \notag \\
  \bar{u}_\pm(k) & = \bar{u}(k) \frac{1 \mp \gamma_5}{2} &
 \bar{v}_\pm(k) & = \bar{v}(k) \frac{1 \pm \gamma_5}{2}\, .
\end{align}
It is further convenient to introduce the following short-hands
\begin{align}
  \label{eq:spinors_short_hands}
  |i^\pm\rangle & \equiv | k_i^\pm \rangle \equiv u_\pm(k_i) =
                  v_\mp(k_i)
&
 \langle i^\pm | & \equiv \langle k_i^\pm | \equiv \bar{u}_\pm(k_i) = \bar{v}_\mp(k_i)
\end{align}
Which allows to define the basic spinor products by
\begin{align}
  \label{eq:spinor_products}
  \langle k_i k_j\rangle & \equiv \langle k_i^- | k_j^+\rangle = \bar{u}_-(k_i)  u_+(k_j),
&
 [ k_i k_j ] & \equiv \langle k_i^+ | k_j^-\rangle = \bar{u}_+(k_i)  u_-(k_j) .
\end{align}
Further note that
\begin{align}
  \label{eq:zero}
  \langle k_i^\pm | k_j^\pm \rangle & =   0  & \langle k_i^\mp | k_i^\pm \rangle & =   0 .
\end{align}
To evaluate such spinor products it is necessary to pick a certain representation of the Dirac $\gamma$ matrices. In the Dirac representation
\begin{align}
  \label{eq:dirac_mat}
  \gamma^0 & =
             \begin{pmatrix}
               {\bf 1} & {\bf 0} \\ {\bf 0} & - {\bf 1}
             \end{pmatrix},
&
\gamma^i &=
           \begin{pmatrix}
             {\bf 0} & \sigma^i \\ - \sigma^i & {\bf 0}
           \end{pmatrix},
&
\gamma^5 & =
           \begin{pmatrix}
             {\bf 0} & {\bf 1} \\ {\bf 1} & {\bf 0}
           \end{pmatrix}\, .
\end{align}
with $\sigma^i$, $i = 1,\ldots, 3$ the Pauli-matrices. Using the convention of Eq.~\eqref{eq:1} to define light-cone momenta,  the massless spinors can be  written as follows,
\begin{align}
  \label{eq:spinor_def_explicit}
  u_+ (k) & = v_- (k) = \frac{1}{2^{1/4}}
            \begin{pmatrix}
              \sqrt{k^+} \\ \sqrt{k^-} e^{i \phi_k}  \\ \sqrt{k^+} \\ \sqrt{k^-} e^{i \phi_k}
            \end{pmatrix}
&
 u_-(k) & = v_+(k) =  \frac{1}{2^{1/4}}
          \begin{pmatrix}
            \sqrt{k^-} e^{-i\phi_k} \\ - \sqrt{k^+} \\ - \sqrt{k^-} e^{-i\phi_k} \\ \sqrt{k^+}
          \end{pmatrix}
\end{align}
with
\begin{align}
  \label{eq:phase}
  e^{ i  \phi_k} & \equiv   
  \frac{k^1  + i k^2}{ \sqrt{{\bm k}^2}} 
= 
  \sqrt{2} \frac{{\bm k}\cdot {\bm \epsilon}}{\sqrt{{\bm k}^2}}, 
& 
 e^{ -i  \phi_k} & =  \sqrt{2} \frac{{\bm k}\cdot {\bm \epsilon}^*}{\sqrt{{\bm k}^2}}, 
&
{\bm \epsilon} & = \frac{1}{\sqrt{2}} (1,  i)
\end{align}
Using these expressions it is possible to obtain explicit formulae for the spinor brackets
\begin{align}
  \label{eq:brackets_explicit}
  \langle k_i k_j\rangle & = \sqrt{2 k_i^- k_j^+} e^{i \phi_{k_i}}  -  \sqrt{2 k_j^- k_i^+} e^{i \phi_{k_j}}
 = 
\sqrt{2 k_i^+ k_j^+} 
    \left(\frac{{\bm k}_i \cdot {\bm \epsilon}}{k_i^+} -  \frac{{\bm k}_j \cdot {\bm \epsilon}}{k_j^+} \right)
\notag \\
 & =  
 ( k_i^+ k_j^+)^{-\frac{1}{2}} \left( k_j^+ | {\bm k}_i| e^{i \phi_{k_i}}    -  k_i^+ |{\bm k}_j| e^{i \phi_{k_j}} \right)
\notag \\
 [ k_i k_j ]  & = -\sqrt{2 k_i^- k_j^+} e^{-i \phi_{k_i}}  +  \sqrt{2 k_j^- k_i^+} e^{-i \phi_{k_j}}
\notag \\
 &= - \sqrt{2 k_i^+ k_j^+} 
    \left(\frac{{\bm k}_i \cdot {\bm \epsilon}^*}{k_i^+} -  \frac{{\bm k}_j \cdot {\bm \epsilon}^*}{k_j^+} \right) 
=
- ( k_i^+ k_j^+)^{-\frac{1}{2}} \left( k_j^+ | {\bm k}_i| e^{-i \phi_{k_i}}    -  k_i^+ |{\bm k}_j| e^{-i \phi_{k_j}} \right)
\end{align}
All of the above versions are equivalent to each other, but prove useful in specific circumstances. 
We further note that
\begin{align}
  \label{eq:stobrack}
  2 p \cdot k & = \langle p k\rangle [k p]\, , 
& [k_ik_i] &  = 0 =\langle  k_i k_ i\rangle\, ,
& \langle k_i k_j\rangle^* &=  - [k_i k_j] \, ,
\end{align}
where the last relation is particularly useful to invert helicities in
an amplitude.  While the above relations are valid in general, they
are particularly useful for the description of high energy factorized
amplitudes where light-cone momenta $k_i^+$ are conserved. In
particular this enables us to deal with explicit expressions of
brackets in the evaluation and leads to a further simplification of
diagrams. For instance, for brackets involving light-cone vectors
$n, \bar{n}$ the following relations hold
\begin{align}
  \label{eq:nnbar_brackets}
  \langle n \bar{n}\rangle & = \sqrt{2}, & 
 \langle n p \rangle & = \sqrt{2 p^+}, \notag \\
  [n \bar{n}]& = -\sqrt{2}, &  [n p] & = - \sqrt{2 p^+}, \notag \\
\langle \bar{n} p\rangle & = -  \sqrt{2}\frac{{\bm p} \cdot {\bm \epsilon}}{ \sqrt{p^+}}, &
[ \bar{n} p ] & =   \sqrt{2}\frac{{\bm p} \cdot {\bm \epsilon}^*}{ \sqrt{p^+}}.
\end{align}
Since the (physical) Hilbert space of a massless vector is isomorphic to the Hilbert space of a massless spinor, it is further possible to express gluon and photon polarization vectors in terms of the above spinors. One has
\begin{align}
  \label{eq:gluon_pol}
  \epsilon_\mu^{(\lambda = +)} (k,n) & \equiv 
+ \frac{\langle k^+ | \gamma_\mu | n^+ \rangle}{\sqrt{2} \langle n^- |
                                       k^+ \rangle}  =
                                       \left(\epsilon_\mu^{(\lambda =
                                       -)} (k,n) \right)^* \, , \notag \\
  \epsilon_\mu^{(\lambda = -)} (k,n) & \equiv  - 
 \frac{\langle k^- | \gamma_\mu | n^- \rangle}{\sqrt{2} \langle n^+ |
                                       k^- \rangle} =
                                       \left(\epsilon_\mu^{(\lambda =
                                       +)} (k,n) \right)^* \, ,
\end{align}
where $k$ denotes the on-shell four-momentum of the boson and $n$ the
gauge vector. Obviously one has $\epsilon \cdot k = 0 = \epsilon \cdot
n$. These polarization vectors obey the following polarization sum
\begin{align}
  \label{eq:polsum}
  \sum_{\lambda = \pm} \epsilon_\mu^{(\lambda)} (k,n)
\left(\epsilon_\mu^{(\lambda)} (k,n) \right)^* &=
- g_{\mu\nu} + \frac{k_\mu n_\nu + n_\mu k_\nu}{k \cdot n} \, .
\end{align}
Using the Fierz identity, 
\begin{align}
  \label{eq:9}
  \langle i^\pm | \gamma^\mu | j^\pm \rangle  \langle k^\pm | \gamma_\mu | l^\pm \rangle & = 2 \langle i^\pm | k^\mp\rangle \langle l^\mp | j^\pm \rangle,
\end{align}
polarization vectors contracted with Dirac matrices can be written as
\begin{align}
  \label{eq:eps_slashed}
  \slashed{\epsilon}^\pm (k, n) & = \frac{\pm \sqrt{2}}{\langle n^\mp | k^\pm \rangle} \left( | k^\mp \rangle \langle  n^\mp | + | n^\pm \rangle  \langle k^\pm |  \right).
\end{align}
While the treatment of the real gluon is straight-forward within this formalism, the case of the virtual gluon is at first special. Since $l^2 = - Q^2 \neq 0$ in general, the photon has two transverse polarizations which are both annihilated by the light-cone vectors $n$, $\bar{n}$, {\it i.e.} one has $\epsilon^\mu_{\lambda=T} \cdot n = 0 = \epsilon^\mu_{\lambda=T} \cdot \bar{n}$. It is therefore straight forward to define transverse polarization vectors within the spinor helicity as
\begin{align}
  \label{eq:photon_transverse}
  \epsilon_\mu^{(\lambda=\pm)}(l) & \equiv  \pm \frac{\langle \bar{n}^\pm | \gamma^\mu | n^\pm \rangle}{\sqrt{2} \langle {n}^\mp | \bar{n}^\pm \rangle},
& \text{with}
&&
 \sum_{\lambda = \pm} \epsilon_\mu^{(\lambda)} (l)
\left(\epsilon_\nu^{(\lambda)} (l) \right)^* &= - g^t_{\mu\nu},
\end{align}
where $g^t_{\mu\nu}$ has only transverse components. The longitudinal polarization vector is given by
\begin{align}
  \label{eq:long_pol1}
   \epsilon_\mu^{(\lambda=L)}(l) & \equiv  \frac{1}{Q} \left[l^+ \bar{n}^\mu + \frac{Q^2}{2 l^+} n^\mu \right].
\end{align}
In the actual calculation we will make use of QED gauge invariance of the scattering amplitude and define
\begin{align}
  \label{eq:eps_barL}
   \bar{\epsilon}_\mu^{(\lambda=L)}(l) & =  \epsilon_\mu^{(\lambda=L)}(l) - \frac{1}{Q} l_\mu = \frac{Q}{l^+} n_\mu.
\end{align}
While adding a term proportional to $l_\mu$ will give a zero
contribution upon contraction with the amplitude, the above form will
facilitate actual calculations.

\subsection{Calculation of amplitudes}
\label{sec:calc}

To evaluate the set of amplitudes Eq.~\eqref{eq:calA1234} we first note that due to the usual spin sums of spinors we have for any on-shell momentum $p^2 = 0$
\begin{align}
  \label{eq:spinsum}
  \slashed{p} & = | p^+ \rangle \langle p^+ | +   | p^- \rangle \langle p^- |,
\end{align}
which allows us to expand contractions of gamma matrices with momenta
$p, q, k$ of the final state particles as well as with the light-cone
vectors $n, \bar{n}$ in terms of spinors. For the momentum of the
initial state off-shell photon we use the representation
in Eq.~\eqref{eq:mom-initial} and apply Eq.~\eqref{eq:spinsum} to each of the
light-cone vectors. For the internal momentum $k_{1,2}$, which are
integrated over and which are therefore intrinsically off-shell, we use
the following decomposition
\begin{align}
  \label{eq:k12_k12bar}
  k_{1,2}^\mu & = \bar{k}_{1,2}^\mu + \frac{k_{1,2}^2}{2 k_{1,2}^+} n^\mu & \text{with} &&
 \bar{k}_{1,2}^\mu  & = 
k_{1,2}^+ \bar{n}^\mu + \frac{{\bm k}_{1,2}^2}{2 k_{1,2}^+} n^\mu + {k}_{1,2;t}^\mu;
\end{align}
since $\bar{k}_{1,2}^2 = 0$, both vectors used to represent $k_{1,2}$
have a spinor representation. While there are many possibilities to
present a massive four vector in terms of two mass-less four-vectors,
the above presentation has two advantages that make it special: a)
since plus-momenta are conserved during interaction with the
background field, the denominators introduced in
Eq.~\eqref{eq:k12_k12bar} can be directly expressed in terms of plus
momenta of external particles b) making use of Eq.~\eqref{eq:zero} and
the fact that Eq.~\eqref{eq:quarkinteraction2} is directly
proportional to $\slashed{n}$, terms proportional to $n^\mu$ will be
set to zero in most cases. To keep our notation more compact, we
further use
\begin{align}
  \label{eq:14}
  l^\mu & = \bar{l}^\mu - \frac{Q^2}{2 l^+} n^\mu
& \text{with} &&
 \bar{l}^\mu & = l^+ \bar{n}^\mu
\end{align}
In the case of gluons a similar simplifications can be achieved using the decomposition of Eq.~\eqref{eq:k12_k12bar}, since
\begin{align}
  \label{eq:dmunu}
   d^{\mu\nu}(k_2) & = - g^{\mu\nu} 
+ 
\frac{\bar{k}_2^\mu n^\nu + \bar{k}_2^\mu n^\nu }{\bar{k}_2 \cdot n} + \frac{n^\mu n^\nu k_2^2}{2 (n \cdot k_2)^2}.
\end{align}
For the current calculation the gluon polarization tensor will be always contracted with an external polarization vector. In this case one finds
\begin{align}
  \label{eq:11}
   d^{\mu\nu}(k_2) \cdot \epsilon^{\lambda, *}_\nu(k) & = -\epsilon_\mu^{\lambda, *}(\bar{k}_2)
\end{align}
where we made use of the polarization sum, Eq.~\eqref{eq:polsum}, and
\begin{align}
  \label{eq:123}
   \epsilon_\nu^{\pm}(\bar{k}_2)  \cdot \epsilon^{\pm, *}_\nu(k)  & =   - \frac{\langle \bar{k}_2^\pm | n^\mp\rangle \langle k^\mp | n^\pm \rangle}{\langle n^\pm | k^\mp\rangle \langle n^\mp | \bar{k}_2^\pm \rangle} =  - 1 
&
  \epsilon_\nu^{\pm }(\bar{k}_2)  \cdot \epsilon^{\mp, *}_\nu(k)  & = 0.
\end{align}
Equipped with this information, evaluation of the numerators of the diagrams Eq.~\eqref{eq:calA1234} is straight forward. While a direct evaluation is possible --  most efficiently done using Computer algebra systems -- the spinor helicity formalism allows to break down the calculation into independent sub-amplitudes, which allows for a very economic evaluation.  In the following we explain this in full detail using diagram 1 and 3 as examples; the case of diagram 2 and 4 follows then directly.

\subsection{Evaluation of  diagrams 1  and 3}
\label{sec:example-1:-diagram}
To put all the relations above to work and get some practice with the
spinor helicity methods, we consider at first the first and third
diagrams given by $i{\cal A}_1$ and $i{\cal A}_3$. We have
\begin{align}
i{\cal A}^a_1 &=  e g \int d^2 {\bm x}_{1} \, d^2 {\bm x}_{2} \,\int {d^4 k_1 \over (2 \pi)^2} \, e^{ i ({\bm k}_1 - {\bm p} - {\bm k}) \cdot  {\bm x}_{1}} \,   
e^{- i ({\bm k}_{1}   + {\bm q})\cdot  {\bm x}_{2}}  \notag \\
&
\delta (p^+ + k^+ - k_1^+) \delta (k_1^+ - l^+ + q^+)
\, \cdot \, 
t^a  V (x_{1,t}) V^\dagger (x_{2,t}) \, \cdot 
 \, {-i \cdot N_1 \over  \, k_1^2 \, (k_1 - l)^2}  \, ,
\end{align}
where $N_1$ is the numerator which further includes the denominator of the external quark line 
\begin{align}
  \label{eq:N1Lone}
  N_1^{\lambda_\gamma; \lambda_q\lambda_{\bar{q}}\lambda_g} & = 
\frac{1}{(p+k)^2} \cdot  \bar{u}_{\lambda_q} (p)  \left(\slashed{\epsilon}^{(\lambda_g)}\right)^*(k)  (\slashed{p}  +  \slashed{k})  \slashed{n} \slashed{k}_1\slashed{\epsilon}^{(\lambda_\gamma)}(l) (\slashed{k}_1     - \slashed{l})\slashed{n} v_{\lambda_{\bar{q}}}  (q) \, .
\end{align}
Here  $\lambda_\gamma = (L, T= \pm), \lambda_q =( \pm) ,\lambda_{\bar{q}} = (\pm), \lambda_g = (\pm)$  encode the possible helicity states of initial and final state particles.  In the case of amplitude 3 we find
\begin{align}
  \label{eq:3rd_diagram}
  i{\cal A}_3^a  &=  e g  \prod_{j=1}^3\int  d^2 {\bm x }_{j}
\int  \frac{d^4 k_1d^4 k_2}{(2 \pi)^5}   
e^{i ({\bm k}_1 - {\bm k}_2 - {\bm p}) \cdot {\bm x}_{1}} \,
 e^{- i ({\bm k}_{1}  + {\bm q})\cdot {\bm x}_{2}}
\, 
e^{ i ({\bm k}_{2} - {\bm k})\cdot {\bm x}_{3}}
\notag \\
& \delta( p^+ - k_1^+ + k_2^+) \, \delta (k^+_1 - l^+ + q^+)\, \delta (k^+_2 - k^+) \, \cdot \,
 V(x_t) t^b V^\dagger (y_t) U(z_t)^{ba}  
\notag \\
&\frac{(-2 k^+)  N_3}{ 
[(k_1 - k_2)^2 + i \epsilon] [k_1^2 + i \epsilon] [k_2^2 + i \epsilon] [(k - l)^2 + i \epsilon]} \, ,
\end{align}
with $N_3$ defined as
\begin{align}
  \label{eq:N3}
  N_3^{\lambda_\gamma; \lambda_q\lambda_{\bar{q}}\lambda_g} & = \bar{u}_{\lambda_q} 
\slashed{n} (\slashed{k}_1 - \slashed{k}_2)  \left(\slashed{\epsilon}^{(\lambda_g)}\right)^*(k_2) \slashed{k}_1  \slashed{\epsilon}^{(\lambda_\gamma)}(l)  (\slashed{k}_1 - \slashed{l})  \slashed{n} v_{\lambda_{\bar{q}}} (q) \, .
\end{align}
It is now possible to factorize both $N_1$ and $N_3$ which in turn will allow for their efficient evaluation. To this end we split $N_{1,3}$ into two, using the spinor representation of $\slashed{k}_1$, which itself relies on the decomposition in Eq.~\eqref{eq:k12_k12bar}. Note that in this decomposition  the term proportional to $\slashed{n}$ gives a zero result for $N_1$. We therefore have\footnote{Recall that the helicity of the anti-quark is due to helicity conservation in massless QCD always opposite to the quark helicity $\lambda_{\bar{q}} = - \lambda_q$}
\begin{align}
  \label{eq:N1N3_fac}
  N_1^{\lambda_\gamma; \lambda_q\lambda_{\bar{q}}\lambda_g} &  = 
Q_{\gamma^* \to q\bar{q}}^{\lambda_\gamma\lambda_q } (\bar{k}_1) 
\cdot 
Q_{q \to q g}^{\lambda_q\lambda_{g}; 1} (\bar{k}_1) \, ,
\notag \\
N_3^{\lambda_\gamma; \lambda_q \lambda_{\bar{q}}\lambda_g} 
&  = 
Q_{\gamma^* \to q\bar{q}}^{\lambda_\gamma\lambda_q} (\bar{k}_1) 
\cdot 
Q_{q \to q g}^{\lambda_q\lambda_{g}; 3} (\bar{k}_1)
+
\frac{k_1^2}{2 k_1^+} \cdot Q_{\gamma^* \to q\bar{q}}^{\lambda_\gamma\lambda_q} (n) 
\cdot 
Q_{q \to q g}^{\lambda_q\lambda_{g}; 3} (n) \, ,
\end{align}
where
\begin{align}
  Q_{\gamma^* \to q\bar{q}}^{\lambda_\gamma\lambda_q} (\bar{k}_1) 
& = 
\bar{u}_{\lambda_q}(\bar{k}_1) \slashed{\epsilon}^{(\lambda_\gamma)}(l)  (\slashed{k}_1 - \slashed{l})  \slashed{n} v_{\lambda_{\bar{q}}} (q) \, ,
\notag \\
Q_{q \to q g}^{\lambda_q\lambda_{g}; 1} (\bar{k}_1) 
&=
   \frac{1}{(p+k)^2} \cdot \bar{u}_{\lambda_q} (p) \left(\slashed{\epsilon}^{(\lambda_g)}\right)^*(k) (\slashed{p} + \slashed{k}) \slashed{n} {u}_{\lambda_q}(\bar{k}_1) \, ,  \notag \\
Q_{q \to q g}^{\lambda_q\lambda_{g}; 3} (\bar{k}_1)
&=
\bar{u}_{\lambda_q} 
\slashed{n} (\slashed{k}_1 - \slashed{k}_2)  \left(\slashed{\epsilon}^{(\lambda_g)}\right)^*(k_2)  {u}_{\lambda_q}(\bar{k}_1) \, .
\end{align}
We start with the splitting $\gamma^* \to q \bar{q}$, with
configuration where the incoming photon has longitudinal polarization
($ \lambda = L$). We find
\begin{align}
  \label{eq:15}
   Q_{\gamma^* \to q\bar{q}}^{L+} (\bar{k}_1)  & =  \frac{Q}{l^+} [\bar{k}_1 n] (\langle n \bar{k}_1\rangle [\bar{k}_1 n] - \langle n \bar{l}\rangle [\bar{l} n]) \langle n q\rangle  \, ,\notag \\
&= 
- 4\frac{Q}{l^+} (p^+ + k^+)^{\frac{1}{2}} (q^+)^{\frac{3}{2}}  \, ,\notag \\
 Q_{\gamma^* \to q\bar{q}}^{L+} (n)  & =  0  \, ,\notag \\
 Q_{\gamma^* \to q\bar{q}}^{L-} (\bar{k}_1)  & =  \left( Q_{\gamma^* \to q\bar{q}}^{L+} (\bar{k}_1)   \right)^* =  Q_{\gamma^* \to q\bar{q}}^{L+} (\bar{k}_1)  \, ,
\end{align}
where me made use of Eq.~\eqref{eq:stobrack} to invert the helicities   in the last expression. For transverse photon polarizations $T = \pm$ we find
\begin{align}
  Q_{\gamma^* \to q\bar{q}}^{++} (\bar{k}_1)  
& =
  \frac{\sqrt{2}}{\langle n \bar{n}\rangle} [\bar{k}_1 \bar{n}]  ( \langle n \bar{k}_1\rangle [\bar{k}_1 n ] - \langle n \bar{l}\rangle [\bar{l} n] ) \langle n q\rangle 
=
4  {\bm k }_1 \cdot {\bm \epsilon}^* \frac{(q^+)^{\frac{3}{2}}}{ (p^+ + k^+)^{\frac{1}{2}}} \, ,
\notag \\
 Q_{\gamma^* \to q\bar{q}}^{+-} (\bar{k}_1)  
& =
\frac{\sqrt{2}}{\langle n \bar{n}\rangle} \langle \bar{k}_1 n\rangle
[\bar{n} \bar{k}_1] \langle \bar{k}_1 n\rangle [nq]
=
-4  {\bm k }_1 \cdot {\bm \epsilon}^* (p^+ + k^+)^{\frac{1}{2}} (q^+)^{\frac{1}{2}}
\, , \notag \\
 Q_{\gamma^* \to q\bar{q}}^{++} (n) & = 4 (q^+)^{\frac{3}{2}} \, ,   \qquad \qquad \qquad \qquad \qquad    Q_{\gamma^* \to q\bar{q}}^{+-} (n)  = 0 \, , \notag \\  
Q_{\gamma^* \to q\bar{q}}^{-+} (\bar{k}_1)  & =
 \left(Q_{\gamma^* \to q\bar{q}}^{+-} (\bar{k}_1)  \right)^* \, ,
 \qquad \qquad \qquad 
 Q_{\gamma^* \to q\bar{q}}^{--} (\bar{k}_1) 
  = \left(Q_{\gamma^* \to q\bar{q}}^{++} (\bar{k}_1)  \right)^* \, .
\end{align}
For the splitting $q \to q g $ it is obviously necessary to distinguish between amplitude one (interaction  before the splitting) and three (interaction after the splitting). For the first case we obtain
\begin{align}
  Q^{++;1}_{q \to qg}
 & = - \frac{\sqrt{2} [pn][pn] \langle n \bar{k}_1\rangle}{[nk] [pk]}
   = 
\frac{2^{\frac{3}{2}} \cdot (p^+)^{\frac{3}{2}} (p^+ + k^+)^{\frac{1}{2}}}{p^+ |{\bm k}|e^{ -i \phi_k} - k^+ |{\bm p}| e^{- i \phi_p}} \, ,
 \notag \\
 Q^{+-;1}_{q \to qg} & = \frac{\sqrt{2} (\langle np\rangle [pn] + \langle nk\rangle  [kn] ) \langle n \bar{k}_1\rangle}{\langle n k\rangle\langle k p\rangle}
=
 \frac{2^{\frac{3}{2}} \cdot (p^+)^{\frac{1}{2}} (p^+ + k^+)^{\frac{3}{2}}}{p^+ |{\bm k}|e^{ i \phi_k} - k^+ |{\bm p}| e^{ i \phi_p}} \, , \notag \\
Q^{- -;1}_{q \to qg} &= \left(Q^{++;1}_{q \to qg} \right)^* \, , \qquad \qquad \qquad 
Q^{- +;1}_{q \to qg} = \left(Q^{+-;1}_{q \to qg} \right)^* \, ,
\end{align}
where we used $(p+ k)^2 = [pk]\langle k p\rangle$. Note that these
splittings allow for a straight forward identification (and if needed
isolation) of collinear and soft singularities. Indeed, these
splittings encode (together with the corresponding
$\bar{q} \to \bar{q}g$ splittings in amplitude 2) the complete soft
($k \to 0$) and collinear ($k \propto p$) singularities of the
process, corresponding to the vanishing of the brackets
$\langle p^\pm | k^\mp \rangle$. The corresponding splitting which
occurs in the third amplitude is on the other hand free of such
potentially singular configurations.  One finds\footnote{Note that in
  slight abuse of notation, the argument $\bar{k}_1$ is meant to apply
  only to the very last spinor in the chain}
\begin{align}
  Q_{q \to qg}^{(++;3)} (\bar{k_1})& = \frac{\sqrt{2} \cdot [pn] (\langle n \bar{k}_1\rangle [\bar{k}_1 n] - \langle n \bar{k}_2\rangle [\bar{k}_2 n]) \langle \bar{k}_2 \bar{k}_1 \rangle}{[n \bar{k}_2]} 
\notag \\
&=  4 (p^+)^{\frac{3}{2}} (p^+ + k^+)^{\frac{1}{2}} \left( \frac{{\bm k}_1 \cdot {\bm \epsilon}}{p^+ + k^+} - \frac{{\bm k}_2 \cdot {\bm \epsilon}}{k^+} \right) \, ,
\notag \\
 Q_{q \to qg}^{(+-;3)} (\bar{k_1})& = - \frac{\sqrt{2}}{\langle n {\bar k}_2\rangle} [pn] \langle n \bar{k}_1\rangle [\bar{k}_1 \bar{k}_2] \langle n \bar{k}_1\rangle \notag \\
& =
 4 (p^+)^{\frac{1}{2}} (p^+
 + k^+)^{\frac{3}{2}} \left( \frac{{\bm k}_1 \cdot {\bm \epsilon}^*}{p^+ + k^+} - \frac{{\bm k}_2 \cdot {\bm \epsilon}^*}{k^+} \right)\notag \\
 Q_{q \to qg}^{(++;3)} (n)& = \frac{\sqrt{2} \cdot [pn] (\langle n \bar{k}_1\rangle [\bar{k}_1 n] - \langle n \bar{k}_2\rangle [\bar{k}_2 n]) \langle \bar{k}_2 n  \rangle}{[n \bar{k}_2]}
= 4 (p^+)^{\frac{3}{2}}  \, , 
\notag \\
 Q_{q \to qg}^{(+-;3)} (n)& =0  \, ,  \qquad \qquad 
Q^{- -;3}_{q \to qg} = \left(Q^{++;3}_{q \to qg} \right)^* \, , \qquad \qquad 
Q^{- +;3}_{q \to qg} = \left(Q^{+-;3}_{q \to qg} \right)^* \, .
\end{align}

\subsection{Diagrams 2 and 4 from  symmetry}
\label{sec:diagrams-2-4}

The corresponding expressions for diagrams 2 and 4 read: 
\begin{align}
i{\cal A}^a_2 &=  e g \int d^2 {\bm x}_{1} \, d^2 {\bm x}_{2} \,\int {d^4 k_1 \over (2 \pi)^2} \, 
e^{- i ({\bm q} + {\bm k} - {\bm k}_{1}) \cdot  {\bm x}_{2}} \,   
e^{- i ({\bm k}_{1}   + {\bm p})\cdot  {\bm x}_{1}}  \notag \\
&
\delta (q^+ + k^+ - k_1^+) \delta (k_1^+ - l^+ + p^+)
\, \cdot \, 
  V (x_{1,t}) V^\dagger (x_{2,t})  t^a \, \cdot 
 \, {-i \cdot N_2 \over  \, k_1^2 \, (k_1 - l)^2} \, ,
\end{align}
where $N_2$ is the numerator which further includes the denominator of
the external anti-quark line
\begin{align}
  \label{eq:N2Lone}
  N_2^{\lambda_\gamma; \lambda_q\lambda_{\bar{q}}\lambda_g} 
& =
  \frac{1}{(q+k)^2} \cdot \bar{u}_{\lambda_q} (p)  \slashed{n} (\slashed{l} - \slashed{k}_1)  \slashed{\epsilon}^{(\lambda_\gamma)}(l) ( - \slashed{k}_1 )   \slashed{n}   
 (-\slashed{q} - \slashed{k})  \left(\slashed{\epsilon}^{(\lambda_g)}\right)^*(k)  v_{\lambda_{\bar{q}}} (q)\, ,
\end{align}
and
\begin{align}
  \label{eq:4th_diagram}
  i{\cal A}_4^a  &=  e g  \prod_{j=1}^3\int  d^2 {\bm x}_{j}
\int  \frac{d^4 k_1d^4 k_2}{(2 \pi)^5}   
e^{- i ({\bm q} - {\bm k}_{1} + {\bm k}_{2}) \cdot {\bm x}_{2}} \,
 e^{- i ({\bm k}_{1}  + {\bm p})\cdot {\bm x}_{1}}\, 
e^{+ i ({\bm k}_{2} - {\bm k})\cdot {\bm x}_{3}}\notag \\
& \delta( q^+ - k_1^+ + k_2^+) \, \delta (k^+_1 - l^+ + p^+)\, \delta (k^+_2 - k^+) \, \cdot \,
 V(x_{1,t}) t^b V^\dagger (x_{2,t}) U(x_{3,t})^{ba}  
\notag \\
&\frac{(-2 k^+)  N_4}{ 
[(k_1 - k_2)^2 + i \epsilon] [k_1^2 + i \epsilon] [k_2^2 + i \epsilon] [(k_1 - l)^2 + i \epsilon]}\,, 
\end{align}
with $N_4$ defined as
\begin{align}
  \label{eq:N4}
  N_4^{\lambda_\gamma; \lambda_q\lambda_{\bar{q}}\lambda_g} & = \bar{u}_{\lambda_q} (p)
    \slashed{n} 
   (\slashed{l} - \slashed{k}_1)  \slashed{\epsilon}^{(\lambda_\gamma)}(l)   \slashed{k}_1  \left(\slashed{\epsilon}^{(\lambda_g)}\right)^*(k_2) (\slashed{k}_2 - \slashed{k}_1)\slashed{n}
v_{\lambda_{\bar{q}}} (q) \, .
\end{align}
Expanding $N_{2,4}$ in terms of spinors one realizes immediately that the
resulting expression is identical to $N_{1,3}$ up to a) interchange of
momenta $p \leftrightarrow q$ b)the order in which spinors are written and
c) a minus sign for each propagator numerator. In particular one finds
the following factorization of $N_2$ and $N_4$:
\begin{align}
  \label{eq:20}
  N_2^{\lambda_\gamma; \lambda_q\lambda_{\bar{q}}\lambda_g} &  =                               - 
\bar{Q}_{\gamma^* \to q\bar{q}}^{\lambda_\gamma\lambda_q} (\bar{k}_1) 
 \cdot 
{Q}_{\bar{q} \to \bar{q} g}^{\lambda_q\lambda_{g}; 2} (\bar{k}_1) \, , \notag \\
  N_4^{\lambda_\gamma; \lambda_q \lambda_{\bar{q}} \lambda_g} &  =                               - 
\bar{Q}_{\gamma^* \to q\bar{q}}^{\lambda_\gamma\lambda_q} (\bar{k}_1) 
 \cdot 
{Q}_{\bar{q} \to \bar{q} g}^{\lambda_q\lambda_{g}; 4} (\bar{k}_1)
-
\bar{Q}_{\gamma^* \to q\bar{q}}^{\lambda_\gamma\lambda_q} (n) 
 \cdot 
{Q}_{\bar{q} \to \bar{q} g}^{\lambda_q\lambda_{g}; 4} (n) \, ,
\end{align}
where 
\begin{align}
  \bar{Q}_{\gamma^* \to q\bar{q}}^{\lambda_\gamma\lambda_q} (\bar{k}_1) 
& = 
\bar{u}_{\lambda_q}(p)  \slashed{n}   (\slashed{k}_1 - \slashed{l}) \slashed{\epsilon}^{(\lambda_\gamma)}(l)   v_{\lambda_{\bar{q}}} (\bar{k}_1) \, ,
\notag \\
Q_{\bar{q} \to \bar{q} g}^{\lambda_q\lambda_{g}; 2} (\bar{k}_1) 
&=
   \frac{1}{(q+k)^2} \cdot \bar{u}_{\lambda_q} (\bar{k}_1) \slashed{n}
 \left(\slashed{\epsilon}^{(\lambda_g)}\right)^*(k) (-\slashed{q} - \slashed{k}) 
 v_{\lambda_{\bar{q}}}(q) \notag \\
Q_{\bar{q} \to \bar{q} g}^{\lambda_q\lambda_{g}; 4} (\bar{k}_1)
&=
\bar{u}_{\lambda_q} (\bar{k}_1) \left(\slashed{\epsilon}^{(\lambda_g)}\right)^*(k_2)
\slashed{n} (\slashed{k}_2 - \slashed{k}_1)   \slashed{n}  v_{\lambda_{\bar{q}}}(q)\, .
\end{align}
Evaluating these expressions one realizes, that they are identical to
the corresponding expressions of diagrams 1 and 3, up to
$q \leftrightarrow p$ and an overall minus sign (due to the different
sign in the numerator of the internal propagators). In case of the
$\bar{q} \to \bar{q} g$ splitting of diagram 2 we find for instance
\begin{align}
  \label{eq:Q2_exp_qbar}
  Q_{\bar{q} \to \bar{q} g}^{-+; 2} (\bar{k}_1) & = \frac{\sqrt{2}}{[nk] [kq]} \langle \bar{k_1} n\rangle
[nq]  [nq] \, .
\end{align}
Using now anti-symmetry of the brackets in the numerator as well as
the bracket $[kq]$ in the denominator, this expression can be
transformed into\footnote{
recall that an anti-quark spinor  with positive (negative) helicity is identical to a quark spinor with negative (positive) helicity
}
\begin{align}
  Q_{\bar{q} \to \bar{q} g}^{(-+; 2)} (\bar{k}_1) 
& = 
\frac{\sqrt{2}  [qn]  [qn]
\langle  n \bar{k_1} \rangle }{[nk] [qk]} 
 = 
-  Q_{q \to q g}^{(++; 1)} (\bar{k}_1 ;   q  \leftrightarrow p)
\, .
\end{align}
We therefore find that the corresponding expressions of diagrams 2 and
4 can be obtained from the corresponding ones of diagram 1 and 3
through a) inverting the quark helicity b) interchanging momenta $p$
and $q$ and c) an overall minus sign. In detail we have
\begin{align}
  \label{eq:21}
\bar{Q}_{\gamma^* \to q \bar{q} }^{(\lambda_\gamma \lambda_q)} (\bar{k}_1)  
& = -
Q_{\gamma^* \to q \bar{q} }^{(\lambda_\gamma (-\lambda_q))} (\bar{k}_1; p \leftrightarrow q) \, ,  
\notag \\ 
   Q_{\bar{q} \to \bar{q} g}^{(\lambda_q \lambda_g; 2)} (\bar{k}_1)  & = 
-  Q_{{q} \to \bar{q} g}^{(-\lambda_q \lambda_g; 1)} (\bar{k}_1; p \leftrightarrow q)
   \, ,\notag \\
 Q_{\bar{q} \to \bar{q} g}^{(\lambda_q \lambda_g; 4)} (\bar{k}_1)  
& =  - Q_{{q} \to \bar{q} g}^{(-\lambda_q \lambda_g; 3)} (\bar{k}_1; p \leftrightarrow q)   \, .
\end{align}

\subsection{Wilson lines and color algebra}
\label{sec:wilson-lines-color}

To this end we recall that the amplitudes in Eq.~\eqref{eq:calA1234} carry both fundamental and adjoint color indices. Making color indices of the fundamental representation explicit, we have $\mathcal{A}^a_{\#, ij}$ where $\# = 1,\ldots, 4$. To write down the differential cross-section we are in general dealing with expressions of the form
\begin{align}
  \label{eq:25}
  & \mathcal{A}^a_{\#_1, ij} \mathcal{A}^{a, \dagger}_{\#_2, ij}, 
&&
& \#_i & = 1, \ldots, 4
\end{align}
with indices $a, i, j$  summed over. It is now convenient to rewrite this as
\begin{align}
  \label{eq:25b}
  \mathcal{A}^a_{\#_1, ij} \mathcal{A}^{a, \dagger}_{\#_2, ij} & =  2 \mathcal{A}^a_{\#_1, ij} t^a_{kl} t^b_{lk} \mathcal{A}^{b, \dagger}_{\#_2, ij} \, ,
\end{align}
which allows us to define amplitudes with four fundamental color
indices only. Extracting factors of Wilson lines and SU$(N_c)$
generators, we find in this way
\begin{align}
  \label{eq:26}
  \left[ V^\dagger(\bx_2)  V(\bx_1)  t^a \right]_{ij} t^a_{kl} & = \frac{1}{2}  \left[V^\dagger(\bx_2) V(\bx_1)  \right]_{il} \delta_{jk} - \frac{1}{2 N_c}  \left[V^\dagger(\bx_2) V(\bx_1)  \right]_{ij}\delta_{kl} \notag \\
\left[  t^a V^\dagger(\bx_2)  V(\bx_1)  \right]_{ij} t^a_{kl} & = \frac{1}{2}  \delta_{il}  \left[V^\dagger(\bx_2) V(\bx_1)  \right]_{jk}  - \frac{1}{2 N_c}  \left[V^\dagger(\bx_2) V(\bx_1)  \right]_{ij}\delta_{kl} \notag \\
 \left[ V^\dagger(\bx_2)  t^b   V(\bx_1) \right]_{ij} U^{ab}(\bx_3)  t^a_{kl} & = 
 \left[ V^\dagger(\bx_2)  V(\bx_3) t^a  V^\dagger(\bx_3) V(\bx_1) \right]_{ij}   t^a_{kl}  \notag \\
& \hspace{-2cm}= \frac{1}{2} \left[ V^\dagger(\bx_2)  V(\bx_3) \right]_{il}  
\left[  V^\dagger(\bx_3) V(\bx_1)\right]_{kj} - \frac{1}{2N_c}  \left[V^\dagger(\bx_2) V(\bx_1)  \right]_{ij}\delta_{kl}
\end{align}
where the first two lines corresponds to diagram one and two
respectively while the last line gives the corresponding factor of
diagrams three and four. To determine the operators of Wilson lines at cross-section level, we restrict at first to the leading $N_c$ terms and subtract
contributions without target interaction (as indicated in
Eq.~\eqref{eq:2}). Extracting an overall factor $ N_c^2/2$  and using the conventional definitions of  dipole and quadrupole 
 \begin{align}
   S^{(2)}_{({\bm x}_1 {\bm x}_2)} 
 & \equiv 
\frac{1}{N_c} \tr \left[ V({\bm x}_1) V^{\dagger}( {\bm x}_2) \right],  
&  S^{(4)}_{({\bm x}_1 {\bm x}_2 {\bm x}_3 {\bm x}_4)}
 &\equiv \frac{1}{N_c} \tr \left[ V({\bm x}_1) V^{\dagger}( {\bm x}_2) V({\bm x}_3) V^{\dagger}( {\bm x}_4) \right],
 \end{align}
we obtain the following set of operators,
\begin{align}
  \label{eq:18large}
 &  N^{(4)}({\bm x}_1,{\bm x}_2,{\bm x}_3,{\bm x}_4)  \equiv  
  1 + S^{(4)}_{({\bm x}_1{\bm x}_2{\bm x}_3{\bm x}_4)}    -  S^{(2)}_{({\bm x}_1{\bm x}_2) }
-  S^{(2)}_{({\bm x}_3{\bm x}_4)} \, ,
\notag \\
 &N^{(22)} ({\bm x}_1,{\bm x}_2 |{\bm x}_3,{\bm x}_4)   \equiv  
 \left[S^{(2)}_{({\bm x}_1 {\bm x}_2)} -1 \right] \left[S^{(2)}_{({\bm x}_3 {\bm x}_4)} -1 \right]
\notag \\
 &N^{(24)}({\bm x}_1,{\bm x}_2|{\bm x}_3,{\bm x}_4,{\bm x}_5,{\bm x}_6) 
 \equiv 
        1  + 
S^{(2)}_{({\bm x}_1 {\bm x}_2)}  S^{(4)}_{({\bm x}_3{\bm x}_4{\bm x}_5{\bm x}_6)}  
   -   S^{(2)}_{({\bm x}_1 {\bm x}_2)}   S^{(2)}_{({\bm x}_3 {\bm x}_6 )}
- 
S^{(2)}_{({\bm x}_4 {\bm x}_5)}
\, ,
\notag \\
  & N^{(44)}({\bm x}_1,{\bm x}_2,{\bm x}_3,{\bm x}_4| {\bm x}_5,{\bm x}_6, {\bm x}_7, {\bm x}_8)    \equiv  
1  + 
 S^{(4)}_{({\bm x}_1{\bm x}_2{\bm x}_3{\bm x}_4) }  S^{(4)}_{({\bm x}_5{\bm x}_6{\bm x}_7{\bm x}_8)   }
\notag \\
& \hspace{7cm}
  - S^{(2)}_{({\bm x}_1{\bm x}_4)} S^{(2)}_{({\bm x}_5 {\bm x}_8)}      - S^{(2)}_{({\bm x}_2 {\bm x}_3) }S^{(2)}_{({\bm x}_6 {\bm x}_7) }
\, ,
\end{align}
 which are used to write down our final result for the differential cross-section Eq.~\eqref{eq:19}. To obtain sub-leading
terms in $N_c$, the operators $N^{(4)}, N^{(22)}, N^{(24)}, N^{(44)}$
need to be replaced by $1/N_c \cdot N^{(4)}({\bm x}_{1,} {\bm x}_{1'},
{\bm x}_{2'}, {\bm x}_2)$. \\

Diffractive reactions (where the color exchange is at
amplitude level restricted to the color singlet) require to project
the $q\bar{q}g$ system onto an overall color-singlet. This leads  to the following replacement of Eq.~\eqref{eq:25}
\begin{align}
  \label{eq:25bdiff}
  \mathcal{A}^a_{\#_1, ij} P^{ab}_{ji;i'j'}  \mathcal{A}^{a,
  \dagger}_{\#_2, i'j'} & = 2 \mathcal{A}^a_{\#_1, ij} t^a_{ji} t^b_{i'j'} \mathcal{A}^{b, \dagger}_{\#_2, i'j'} \, .
\end{align}
At the level of Eq.~\eqref{eq:26} this corresponds to replacing
$t^a_{kl}$ by $t^a_{ji}$ on the right-hand side which translates into
contracting the left-hand side with $\delta_{jk}\delta_{li}$. As a
results one finds that all quadrupole operators in
Eq.~\eqref{eq:18large} factorize into a product of dipole
operators. One finds for the leading $N_c$ terms:
\begin{align}
  \label{eq:18large_diff}
 &  N^{(4)}_{\text{diff.}}({\bm x}_1,{\bm x}_2,{\bm x}_3,{\bm x}_4)  \equiv  
   \left[  1  -  S^{(2)}_{({\bm x}_1{\bm x}_2) } \right] \left[
1-  S^{(2)}_{({\bm x}_3{\bm x}_4)} \right] \, ,
\notag \\
 &N^{(22)}_{\text{diff.}} ({\bm x}_1,{\bm x}_2 |{\bm x}_3,{\bm x}_4)   \equiv  
 \left[S^{(2)}_{({\bm x}_1 {\bm x}_2)} -1 \right] \left[S^{(2)}_{({\bm x}_3 {\bm x}_4)} -1 \right]\, ,
\notag \\
 &N^{(24)}_{\text{diff.}}({\bm x}_1,{\bm x}_2|{\bm x}_3,{\bm x}_4,{\bm x}_5,{\bm x}_6) 
 \equiv 
\left[         1  - 
S^{(2)}_{({\bm x}_1 {\bm x}_2)}  S^{(2)}_{({\bm x}_3{\bm x}_6)}   \right] \left[ 1 
- 
S^{(2)}_{({\bm x}_4 {\bm x}_5)} \right]
\, ,
\notag \\
  & N^{(44)}_{\text{diff.}}({\bm x}_1,{\bm x}_2,{\bm x}_3,{\bm x}_4| {\bm x}_5,{\bm x}_6, {\bm x}_7, {\bm x}_8)    \equiv   \left[ 
1  
  - S^{(2)}_{({\bm x}_1{\bm x}_4)} S^{(2)}_{({\bm x}_5 {\bm x}_8)}  \right] 
\left[ 1     - S^{(2)}_{({\bm x}_2 {\bm x}_3) }S^{(2)}_{({\bm x}_6 {\bm x}_7) } \right]
\, .
\end{align}
To obtain sub-leading
terms in $N_c$, it is again necessary to replace the operators $N_{\text{diff}}^{(4)}$, $N_{\text{diff}}^{(22)}$, $N_{\text{diff}}^{(24)}$, $N_{\text{diff}}^{(44)}$
 by $1/N_c \cdot N_{\text{diff}}^{(4)}({\bm x}_{1,} {\bm x}_{1'},
{\bm x}_{2'}, {\bm x}_2)$.

 \subsection{Integrals}
\label{sec:integrals}

Integrations over plus-momenta are carried out trivially using the delta-functions associated with the vertices Eqs.~\eqref{eq:quarkinteraction2} and ~\eqref{eq:gluoninteraction2}. Integrations over minus-momenta can be performed using contour integrations. In the case of diagram one and two one  finds
\begin{align}
  \label{eq:22}
I_1 & =   -i \int {d k_1^+ \over 2 \pi}  \int {d k_1^- \over 2 \pi} \,
\frac{(2 \pi)^2 \delta(p^+ - k^+ - k_1^+) \delta(k_1^+ - l^+ - q^+)}{ [k_1^2 + i \epsilon]  [(k_1 - l)^2 + i \epsilon}
\notag \\
&=
2 \pi \delta (l^+ - p^+ - k^+ - q^+) \frac{1}{2 l^+} {1 \over \left[ k_{1 t}^2 + z_2 (z_1 + z_3) Q^2\right]} \notag \\
I_2 & = I_1 \left(\left\{ p^+ \leftrightarrow q^+ \right\} \right)
\end{align}
with photon momentum fractions $z_1 = p^+/l^+$, $z_2 = q^+/l^+$ and
$z_3 = k^+/l^+$. In the case of diagrams three and four two (closely related) integrals
are needed for each diagram. One has
\begin{align}
  \label{eq:23}
  I_{3,1} & = 
\int \frac{d k_1^+}{2 \pi}\int \frac{d k_2^+}{2 \pi}
\int \frac{d k_1^-}{2 \pi}\int \frac{d k_2^-}{2 \pi}
(2 \pi)^3 \delta(p^+ - k_1^+ + k_2^+) \delta(k_1^+ - l^+ + q^+)
        \delta(k_2^+ - k^+) \notag \\
& \hspace{4cm}
\cdot \frac{ (-2 k^+)}
{[(k_1 - k_2)^2 + i \epsilon] [k_1^2 + i \epsilon] [k_2^2 + i \epsilon] [(k - l)^2 + i \epsilon]}
\notag \\
& =
2 \pi \delta(l^+ - p^+ - k^+ - q^+) \frac{1}{ 2 z_1  l^+}
\frac{1}{[z_2(1-z_2) Q^2 + {\bm k}_1^2] \left[Q^2 + \frac{{\bm k}_1^2}{z_2} + \frac{{\bm k}_2^2}{z_3} + \frac{({\bm k}_1 - {\bm k}_2)^2}{z_1}\right]}
\notag \\
 I_{3,2} & = 
\int \frac{d k_1^+}{2 \pi}\int \frac{d k_2^+}{2 \pi}
\int \frac{d k_1^-}{2 \pi}\int \frac{d k_2^-}{2 \pi}
(2 \pi)^3 \delta(p^+ - k_1^+ + k_2^+) \delta(k_1^+ - l^+ + q^+)
        \delta(k_2^+ - k^+) \notag \\
& \hspace{4cm}
\frac{1}{2 k_1^+} \cdot \frac{ (-2 k^+)}
{[(k_1 - k_2)^2 + i \epsilon]  [k_2^2 + i \epsilon] [(k - l)^2 + i \epsilon]}
\notag \\
& =
2 \pi \delta(l^+ - p^+ - k^+ - q^+) \frac{-1}{4 z_1 z_2 (1-z_2)  \cdot (l^+)^2}
\frac{1}{ \left[Q^2 + \frac{{\bm k}_1^2}{z_2} + \frac{{\bm k}_2^2}{z_3} + \frac{({\bm k}_1 - {\bm k}_2)^2}{z_1}\right]}
\notag \\
I_{4,i} & = I_{3,i} \left(\left\{ p^+ \leftrightarrow q^+ \right\}
          \right), \qquad \qquad i = 1,2.
\end{align}
If one wishes to evaluate Eq.~\eqref{eq:2} with gluon correlators in
the target given in transverse momentum space, the above form of the
integrals is already sufficient. If gluon correlators in the target
are provided in transverse coordinate space, the corresponding Fourier
transforms can be be carried out using the integrals of appendix B of
\cite{Beuf:2011xd}, which then allow to write our result in the large
$N_c$ limit in the form already presented in \cite{ahjt-1}. For completeness we present the large $N_c$ result below\footnote{Note that we corrected an error present in Eq.~(12) of \cite{ahjt-1} as well as the a missing average factor for the transverse cross-section}. 
\subsection{The (large $N_c$) result}
\label{sec:large-n_c-result}

With $\alpha_{em}$ and $\alpha_s$ the electromagnetic and strong
coupling constants, and $e_f$ the electro-magnetic charge of the quark
with flavor $f$ we obtain the following leading $N_c$
result\footnote{We note that the result presented below slightly
  differs from the  result reported in the letter   \cite{ahjt-1}
  where  an erroneous overall factor of $1/(2 \pi)^2$ has been
  included; we further corrected typos present in the expressions
  corresponding to  Eq.~\eqref{eq:M_L} and Eq.~\eqref{eq:a1L}  in \cite{ahjt-1}.}:
\begin{align}
  \label{eq:19}
&   \frac{d \sigma^{T,L}}{d^2{\bm p}\, d^2 {\bm k} \, d^2 {\bm q} \, dz_1 dz_2}    = \notag \\
& \quad  = c_{T,L} \cdot \frac{\alpha_s   \alpha_{em} e_f^2 N_c^2 }{z_1 z_2 z_3 \cdot  2 } 
  \prod_{i=1}^3 \prod_{j=1}^3   \int \frac{d^2 {\bm x}_i}{(2 \pi)^2}
 \int   \frac{d^2 {\bm x}_j'}{(2 \pi)^2}   e^{i {\bm p}({\bx_1 - \bx_1'}) + i {\bm q}({\bx_2 - \bx_2'}) + i {\bm k}({\bx_3 - \bx_3'})}
  \notag 
\\ & \qquad  \bigg\langle
(2 \pi)^4\bigg[ \bigg( 
 \delta^{(2)}(\bx_{13})\delta^{(2)}(\bx_{1'3'})\sum_{h, g}  \psi^{T,L}_{1;h,g}({\bm x}_{12})   \psi^{T,L,*}_{1';h,g}({\bm x}_{1'2'}) 
+
\{ 1, 1' \}  \leftrightarrow \{ 2, 2'\}
\bigg)  \notag \\
& 
 \quad  \qquad \cdot   N^{(4)}({\bm x}_1,{\bm x}_1',{\bm x}_2',{\bm x}_2)
+
\bigg( 
 \delta^{(2)}(\bx_{23})\delta^{(2)}(\bx_{1'3'})\sum_{h, g}  \psi^{T,L}_{2;h,g}({\bm x}_{12})   \psi^{T,L,*}_{1';h,g}({\bm x}_{1'2'}) 
\notag \\
& \hspace{6.6cm}
+
\{ 1, 1' \}  \leftrightarrow \{ 2, 2'\}
\bigg)\cdot
 N^{(22)}({\bm x}_1,{\bm x}_1'|{\bm x}_2',{\bm x}_2)
 \bigg]\notag \\
& \quad 
 + (2 \pi)^2\bigg[
\delta^{(2)}(\bx_{13}) \sum_{h, g}  \psi^{T,L}_{1;h,g}({\bm x}_{12})   \psi^{T,L,*}_{3';h,g}({\bm x}_{1'3'}, \bx_{2'3'}) N^{(24)}(\bx_{3'}, \bx_{1'}| \bx_{2'} ,\bx_2 ,\bx_1 ,\bx_{3'}) 
 \notag \\
&  \quad \qquad   +
\{ 1 \}  \leftrightarrow \{ 2\}
+
 \delta^{(2)}(\bx_{1'3'}) \sum_{h, g}  \psi^{T,L}_{3;h,g}({\bm x}_{13}, \bx_{23})   \psi^{T,L,*}_{1';h,g}({\bm x}_{1'2'})
\notag \\
& \hspace{6.4cm} \cdot 
 N^{(24)}(\bx_{1}, \bx_{3}| \bx_{2'} ,\bx_2 ,\bx_3 ,\bx_{1'}) 
 +
\{ 1' \}  \leftrightarrow \{ 2'\}
\bigg]
\notag \\
& \quad 
+ \sum_{h, g}  \psi^{T,L}_{3;h,g}({\bm x}_{13}, \bx_{23})   \psi^{T,L,*}_{3';h,g}({\bm x}_{1'3'}, \bx_{2'3'}) \cdot N^{(44)}(\bx_{1}, \bx_{1'}, \bx_{3'}, \bx_3| \bx_{3} ,\bx_{3'} ,\bx_{2'} ,\bx_{2}) 
 \bigg\rangle_{A^+} \, ,
\end{align}  
where $\psi_{i'} \equiv \psi_{i}$, $i=1, \ldots, 3$ and
$z_3 = 1-z_1 - z_2$, while $c_L =1$, $c_T= 1/2$. To obtain sub-leading terms in $N_c$ all
operators $N^{(4)}, N^{(22)}, N^{(24)}, N^{(44)}$ are to be replaced
by $1/N_c \cdot N^{(4)} (\bx_1, \bx_{1'}, \bx_{2'}, \bx_2)$.  With
$ \phi_{{ij}}$ the azimuthal angle of ${\bx}_{ij}$, $i,j=1\ldots, 3$
and
\begin{align}
X_j^2 & = \bx_{12}^2 (z_j + z_3)
   \left(1-z_j - z_3\right), \quad j=1,2,  & 
  X_3^2 & = z_1 z_2 \bx_{12}^2+  z_1 z_3 \bx_{13}^2+ z_2 z_3 \bx_{23}^2 \, ,
\end{align} we obtain
  \begin{align}
  \label{eq:M_L}
  \psi_{j,hg}^{L}  &= -2\sqrt{2}  Q  
 K_0\left(Q X_j\right) \cdot a_{j,hg}^{(L)},   &   j &=1,2 
\notag \\
  \psi_{j,hg}^{T}  &=  2 i e^{\mp i \phi _{\bx_{12}}}
{  \sqrt{(1-z_3 - z_j) (z_j + z_3)}Q   K_1\left(Q X_j \right) }
\cdot   a_{j,hg}^{\pm}  &   j &=1,2  \notag \\
 \psi_{3,hg}^{L}  &= 
4  \pi i Q  \sqrt{2 z_1 z_2}    K_0\left(Q  X_3 \right)  (a_{3,hg}^{(L)} + a_{4,hg}^{(L)}), \notag \\ 
 \psi_{3,hg}^{T}  &= -
4  \pi Q  \sqrt{ z_1 z_2}    \frac{K_1\left(Q  X_3 \right)}{X_3}  (a_{3,hg}^{\pm} + a_{4,hg}^{\pm})\, .
\end{align}
With
\begin{align}
 & a_{k+1,hg}^{T,L}  =  -  a_{k,-hg}^{T,L} (\{p, \bx_1\} \leftrightarrow \{q, \bx_2\}),\qquad  k = 1,2\, , &
  a_{j,hg}^{T,L}  &= a_{j,-h-g}^{(-T,L)*}, \quad   j=1,\ldots ,4.
\end{align}
we have for longitudinal photon polarizations
\begin{align}
 a^{(L)}_{1,++}&= 
- \frac{(z_1 z_2)^{3/2}  \left(z_1+z_3\right)}{z_3 e^{-i \theta _p} |{\bm p}|-z_1
   e^{-i \theta _k} |{\bm k}|},
 &  a^{(L)}_{1,-+} & = 
-\frac{\sqrt{z_1} z_2^{3/2} \left(z_1+z_3\right){}^2}{z_3 e^{-i \theta _p} |{\bm p}|-z_1
   e^{-i \theta _k} |{\bm k}|},
\notag \\
 a^{(L)}_{3,++} &= \frac{z_1z_2}{|{\bm x}_{13}|e^{-i \phi_{\bx_{13}}}}, &
a^{(L)}_{3,-+}& = \frac{z_2(1-z_2)}{|{\bm x}_{13}|e^{-i \phi_{\bx_{13}}}},
\end{align}
while transverse polarization read
\begin{align}
   a^{(+)}_{1,++}&= -\frac{ (z_1 z_2)^{3/2}}{z_3 e^{-i \theta _p} |{\bm p}|-z_1
   e^{-i \theta _k} |{\bm k}|}, 
 \notag \\
 a^{(+)}_{1,+-}& =
\frac{\sqrt{z_1}(z_2)^{\frac{3}{2}} (z_1 + z_3)}{z_1 e^{i \theta _k} |{\bm k}|-z_3
   e^{i \theta _p} |{\bm p}|} ,
\notag \\
a^{(+)}_{1,-+} & =  \frac{\sqrt{z_1z_2} (z_1 + z_3)^2}
{z_3 e^{-i \theta _p} |{\bm p}|-z_1
   e^{-i \theta _k} |{\bm k}|},
\notag \\
 a^{(+)}_{1,--}&=  \frac{z_1^{3/2} \sqrt{z_2} (z_1 + z_3) }{z_3 e^{i \theta _p} |{\bm p}|-z_1 
   e^{i \theta _k} |{\bm k}|},
\notag \\
a^{(+)}_{3,++} & = \frac{z_1 z_2 (z_2 z_3 |\bx_{23}| e^{- i \phi_{\bx_{23}}} + 
z_3 |\bx_{13}| e^{- i \phi_{\bx_{13}}} - z_1 z_2 |\bx_{12}| e^{- i \phi_{\bx_{12}}} )}{(z_1 + z_3)|\bx_{13}| e^{- i \phi_{\bx_{13}}}},
\notag \\
a^{(+)}_{3,+-} & =  \frac{ z_2^2 (z_3 |\bx_{23}| e^{- i \phi_{\bx_{23}}} - z_1 |\bx_{12}| e^{- i \phi_{\bx_{12}}} )}{|\bx_{13}| e^{ i \phi_{\bx_{13}}}},
\notag \\
a^{(+)}_{3,-+} &=- \frac{z_2(z_1 + z_3) (z_3 |\bx_{23}| e^{- i \phi_{\bx_{23}}} - z_1 |\bx_{12}| e^{- i \phi_{\bx_{12}}} )}{|\bx_{13}| e^{- i \phi_{\bx_{13}}}},
\notag \\
a^{(+)}_{3,--} & = 
\frac{ z_1z_2 ( z_1 |\bx_{12}| e^{- i \phi_{\bx_{12}}} - z_3 |\bx_{23}| e^{- i \phi_{\bx_{23}}}  )}{|\bx_{13}| e^{ i \phi_{\bx_{13}}}}.
 \label{eq:a1L}
\end{align}
These expressions were already used to study azimuthal angular correlations between the three produced partons in DIS where it was shown that gluon saturation effects lead to a broadening and disappearance of the away side peaks. This is qualitatively similar to the disappearance of di-hadron angular correlations DIS~\cite{dis-double} and in the forward rapidity region of high energy proton (deuteron)-nucleus collisions~\cite{cm-double}.

\section{Summary}
\label{sec:summary}
We have derived the triple differential cross section for production of a quark, anti-quark and a gluon in DIS for both transversely and longitudinally polarized photons. The final expression was already published in a short letter~\cite{ahjt-1}, here we show the full details of the calculation. After a discussion of the contributing diagrams in coordinate and momentum spaces, we give a brief overview of spinor helicity techniques and apply it to the process considered which leads to an enormous simplification of the Dirac Algebra involved. Besides being used for studying the effects of gluon saturation dynamics on azimuthal angular correlations of produced hadrons/jets in DIS, the resulting expressions can also be used, with trivial modification, to study three-jet production in ultra-peripheral heavy ion collisions at RHIC and the LHC using the CGC formalism. Furthermore, using the crossing symmetry of the amplitudes one can relate this process to Multi Parton Interactions (MPI) in a proton at large $x$ in processes where one produces a (real or virtual) photon and a jet (in case of DPI) or a (real or virtual) photon in case of TPI in the forward rapidity kinematics. If one assumes that the target is accurately described by the CGC formalism, one can then extract valuable information on intrinsic parton correlations at large $x$ in a proton~\cite{Kovner:2017vro}.

\section*{Acknowledgments}
J.J-M. acknowledges support by the DOE Office of Nuclear Physics
through Grant No.\ DE-FG02-09ER41620 and by the Idex Paris-Saclay though a Jean d'Alembert grant and would like to thank L. Dixon for helpful correspondence on spinor helicity methods. Support has been received in part  by UNAM-DGAPA-PAPIIT grant number IN101515 and by Consejo Nacional de Ciencia y Tecnolog\'{\i}a grant number 256494. M.H. acknowledges support by CONACyT-Mexico grant numbers
CB-2014-22117 and Proy.~No. 241408 as well as the Red-FAE. M.E.T.-Y. 
acknowledges support from CONACyT-M\'exico
sabbatical grant number 232946 and the kind hospitality of ICN-UNAM
provided in the initial stages of this collaboration.

\appendix

\end{document}